\documentclass[sigplan,10pt,nonacm]{acmart}
\renewcommand\footnotetextcopyrightpermission[1]{}

\usepackage{my-style}

\usepackage{tikz}
\usepackage{amsmath}
\usepackage{graphicx}
\usepackage{mathtools}
\usepackage{subcaption}

\usepackage{filecontents}

\AtBeginDocument{%
  }

\setcopyright{none}

\author{Seyed Armin Vakil Ghahani}
\affiliation{%
  \institution{University of Michigan}
  \country{}
}
\email{arminvak@umich.edu}
\author{Manos Kapritsos}
\affiliation{%
  \institution{University of Michigan}
  \country{}
}
\email{manosk@umich.edu}

\begin{document}

\title{Synthesizing Proofs Using Proof Sharding and Exploration}

\renewcommand{\shortauthors}{}

\begin{abstract}
Distributed systems are hard to implement correctly, and subtle bugs can go undetected using traditional testing.
Formal verification offers an alternative for proving the correctness of complex distributed systems.
Despite previous efforts to automate and facilitate formal verification, it is still hard to integrate formal verification in software development.
Programmers need to query the theorem prover repeatedly to find the correct proof of their system.
This cycle of going back and forth with the theorem prover involves a lot of human intervention and is a barrier to adopting formal verification in software development.

In this paper, we address the challenges of scaling formal verification in practice by reducing the human intervention required to find the correctness proof of a distributed system.
We propose \sys, an automated tool that shards large verification tasks and explores the proof search space for each shard independently to synthesize correctness proofs.
We use controlled exploration when adding possible proof annotations to handle the search-space explosion problem.
We evaluate \sys on a variety of safety proofs for distributed systems.
We show that \sys can find the last proof annotation in 86 out of 103 proof-completion tasks, with runtimes ranging from one minute to two hours.
% safety proof of complex distributed systems such as Paxos in 24 minutes, on average.
% This work lays the foundations of how we can scale proof synthesis of complex protocols correctness.
% This paper introduces \ss, a novel approach to automate proof generation by addressing two critical challenges: (1) decomposing complex lemmas into smaller, independently verifiable sub-lemmas, and (2) generating triggers and witnesses to guide SMT solvers toward successful proof completion. \ss enables concurrent verification of sub-lemmas and tailors solver guidance to each proof component, significantly reducing manual effort and improving scalability.

% We evaluate \ss on a leader election example, demonstrating its ability to complete the proof in three minutes. By bridging key gaps in automated formal verification, \ss advances the state of the art, providing developers with an efficient and scalable solution for verifying complex distributed systems.
\end{abstract}

\settopmatter{printfolios=true}
\maketitle
\pagestyle{plain}

\section{Introduction}
% \bubble{Distributed systems are complex and error-prone.}
% Distributed systems are inherently complex and prone to subtle errors due to their complexity.
% Node failures, delayed or reordered messages and network partitions can lead to inconsistent states that are difficult to reason about or reproduce in testing environments.
% These intricacies make distributed systems hard to test using traditional testing allowing critical bugs to go undetected. As a result, these bugs often compromise system availability and impose high costs on users.

\bubble{Formal verification is an alternative to traditional testing to write bug-free programs.}
Distributed systems are notoriously complex and prone to subtle errors.
Despite extensive testing, bugs often remain undetected in production, compromising system availability and imposing high costs on users~\cite{CRNBugs2025,BeschastnikhWBE2016,Yuan2020Lu}.
% \manos{need citations here.}
Formal verification provides an alternative to traditional testing, allowing programmers to reason formally about the correctness of distributed systems and offering strong guarantees about the absence of bugs~\cite{Hawblitzel2015Howell, Wilcox2015Woos, Bornholt2021Joshi, Krogh2020Timany, Newcombe2015Rath, Sharma2023Jung, Hance2020Lattuada, Lattuada2023Hance}.
%Instead of relying on test cases, formal verification uses theorem provers to prove correctness and provides strong guarantees on absence of bugs.
% to ensure safety and liveness properties are hold in every execution. 
% Formal verification provides strong guarantees about the absence of bugs but requires programmers to write formal proofs on correctness.

\bubble{Adopting formal verification to software development is challenging due to its high required manual effort.}
Unfortunately, even with the strong guarantees formal verification provides, adopting it in software development remains costly.
Formal verification demands substantial effort from programmers.
In addition to designing the protocol, programmers need to define a safety property and then write proofs to guarantee that the protocol is safe in all reachable states.
These proofs are usually written as lemmas using Hoare logic~\cite{Hoare1969} and mechanically checked using a theorem prover such as Rocq~\cite{coq}, Dafny~\cite{leino2010dafny}, or Verus~\cite{lattuada2024verus}.

%This process is time-consuming as programmers need to determine what proof annotations are required to convince the theorem prover that their program is correct.

% combine next two. also mention what is the manual effort that programmers need to write.

% \bubble{After solver fails to prove correctness, programmer needs to find the next proof to send to solver manually.}
%Automated theorem provers (e.g. Dafny and Verus~\cite{leino2010dafny, lattuada2024verus}) and prior work~\cite{ma2019i4, Zhang2024Hance, Yao2021Tao, Yao2022Tao, Zhang2025Singh}\manos{What prior work is this? I don't see any references to it later in the paragraph.}\armin{I meant the prior work on automating finding inductive invariants.} significantly reduce the manual effort required for formal verification compared to interactive theorem provers (e.g. Rocq \cite{rocq}). That said, even when the system is bug-free, most lemmas still require proof annotations for the prover to verify correctness. And naturally, the more complex the system, the more proof annotations are required. 

\begin{figure}[t!]
\centering
\includegraphics[width=0.7\columnwidth]{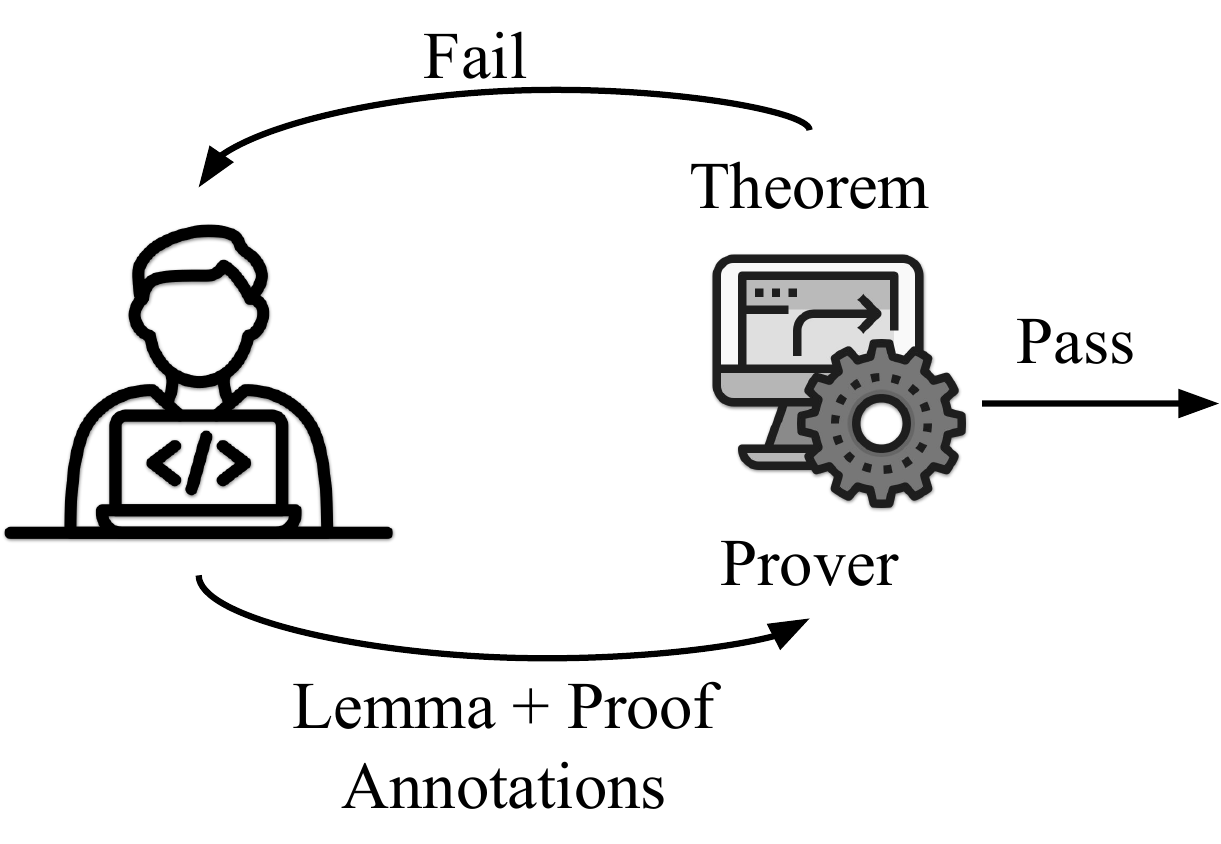}
\caption{The developer needs to query the theorem prover repeatedly to debug the proof and find the sufficient proof annotations for the proof to pass.}
\Description{The developer needs to query the theorem prover repeatedly to debug the proof and find the sufficient proof annotations for the proof to pass.}
\label{fig:proofDebugging}
\end{figure}

\bubble{Automated solvers and prior work help automating verification process but programmers still need to write proofs manually after solvers fail to prove a lemma even if it is bug-free.}
Even though automated provers like Dafny or Verus provide a high level of automation to check these proofs, programmers usually need to add proof annotations to guide the prover and complete the proof.
As a result, programmers spend the majority of their time in a ``proof-debugging'' loop, as shown in Figure~\ref{fig:proofDebugging}:
They send a candidate proof of their lemma to the theorem prover; if the prover fails to verify the lemma using this candidate proof, they have to manually determine what proof annotations to add next, and then try again.
Crucially, theorem provers do not guide the developer on what is the next step to pursue in finding the correct proof.

Cho~et~al. propose ProofPlumber~\cite{Cho2024Zhou} that provides automated tools in code editors to add common proof annotations.
Although this helps programmers generate proof annotations automatically, it is still up to the developer to decide which proof annotation they should try first.
As a result, determining the best step forward still requires programmer expertise.
In other words, ProofPlumber only helps with the bottom arrow in Figure~\ref{fig:proofDebugging} and still requires human input in each iteration with the theorem prover.

% This continuous back and forth between the prover and the developer is one of the main barriers to the practical adoption of formal verification.

% \armin{second, third and fourth paragraph need to be less verbose.}
% \bubble{The main challenge on improving automation of formal verification lies within the way theorem provers are used.}
% The main obstacle of improving the automation of formal verification lies in the way that programmers can currently use theorem provers. 
% In each iteration, they ask the theorem prover if the current program-proof pair is provably correct or not. 
% Using the theorem prover output, they can only gain insights on the current proof.
% Theorem provers do not guide the developer on what is the next step for them to pursue in finding the correct proof and making progress.
% Thus, even if the proof needs minor modifications to pass, a human input is required at all times to modify the failing proof and querying the theorem prover again and again until they find the proof.
% This cycle of going back and forth to the theorem prover and the necessary manual input is labor-intensive and a limiting factor on the scalability and usability of formal verification.

% Current tools do not allow programmers to automatically query the theorem prover to check multiple proofs at the same time for a lemma.

\begin{figure}[t!]
\centering
\includegraphics[width=\columnwidth]{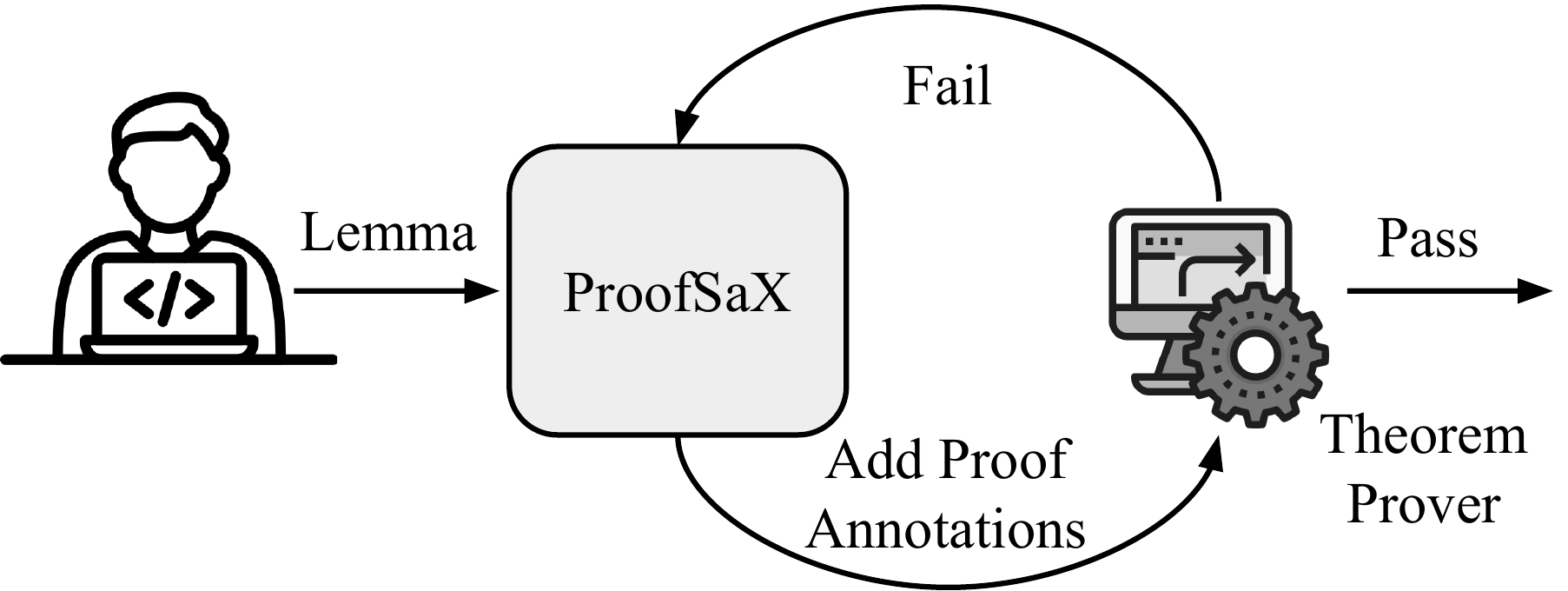}
\caption{The developer invokes \sys to query the theorem prover repeatedly and add proof annotations where necessary to find the complete proof.
}
\Description{The developer invokes \sys to query the theorem prover repeatedly and add proof annotations where necessary to find the complete proof.
}
\label{fig:sysDebugging}
\end{figure}

\bubble{The goal of this project is to automate finding proofs at the program level and complement theorem provers.}
% \bubble{talk about the trade of human time to compute time.}
In this work, our goal is to provide a framework that complements existing theorem provers by automatically generating proof annotations that can guide the theorem prover to find the proof without human intervention, as shown in Figure~\ref{fig:sysDebugging}.
% The goal of this work is to automatically generate the next proof candidates that can be checked by the theorem prover after a failure.
By automating the iterative process of asking the theorem prover whether a particular program-proof pair is correct, we reduce the amount of manual input required to convince the theorem prover of correctness.
% Given a failing lemma, this work adds a variety of proof annotations to this lemma and checks these candidate proofs using the theorem prover.
% This design allows programmers to trade human time with compute time, utilizing a cluster of nodes to enhance the development of formally verified programs.

% \bubble{Although, it is thought that this task can only be done using human intuition, this task is inherently an exhaustive search of the program state, and as a result, can be automated.}
\bubble{In theory, fully automating finding the next proof to check is impossible as this is an undecidable problem. However, just like SAT solving that deemed to be impossible to automate, our goal is automate finding the next proof to check in practice without leading to a search space explosion.}
Fully automating proof search for a lemma in all cases is theoretically impossible since this is an undecidable problem.
However, worst-case impossibility does not prevent automation in practice.
SAT solving, for example, was once viewed as infeasible to automate, yet advances in this area have made fully automated SAT solving practical in various areas of computer science~\cite{Fichte2023Berre, applicationsOfSATInHardware}. 
Similarly, our goal is not to find every proof automatically, but rather to help programmers complete proofs that theorem provers fail to prove automatically without programmer guidance.
We aim to address the practical challenges of scaling the automation of formal verification.

% Note that this is a challenging task\manos{Never say "this is hard". In fact, that's not exactly the point we want to make here; what comes next is the point.} as fully automating finding the proof of a lemma in all cases is theoretically impossible since this is an undecidable problem.
% However, just like SAT solving that was deemed to be impossible to automate\manos{was it really "impossible to automate"?} but is currently adopted in various areas of computer science~\cite{Fichte2023Berre, applicationsOfSATInHardware}, our goal is to facilitate finding the proof of failing lemmas and address the challenges of scaling formal verification.
% We aim to scale the automation of formal verification by focusing on practical heuristics, bounded exploration and parallelization.

% \sys consists of three main components which we will describe in the following paragraphs.

One of the key techniques that system developers use to solve a large task in a practical and scalable setting is to shard the task and process each shard independently.
MapReduce~\cite{Dean2008Ghemawat}, for instance, uses sharding to break large, computationally intensive tasks into smaller tasks. Spanner~\cite{Corbett2013Dean}, Bigtable~\cite{Chang2008Dean}, and Dynamo~\cite{DeCandia2007Hastorun} manage storage at scale using sharding techniques.
We argue that verification is no different.
Verification tasks can be broken into smaller shards and the proof for each shard can be found separately.

\bubble{We present \sys to tackle formal verification scalability.}
In this paper, we present \sys (Proof Shard \& Explore), a technique that divides verification tasks into smaller shards (\textit{proof sharding}) and then automatically explores the proof annotations needed to complete the proof of each shard (\textit{proof exploration}).
\sys uses a divide-and-conquer workflow: it decomposes a lemma into smaller proof obligations, queries the theorem prover on each one, and iterates only on the obligations that the theorem prover cannot prove on its own.
% \sys then concentrates exclusively on these failing cases.

% We present Controlled Proof Synthesis (\sys) to automate finding proof of lemmas where theorem provers fail to prove automatically\manos{weird phrasing}.
% \bubble{Generating proofs blindly is not scalable.}
% \sys employs a Divide-and-Conquer approach to first break down the proof to smaller pieces to determine the point of failure using theorem prover response.
% Next, \sys focuses on cases where the theorem prover fails to prove automatically and find the necessary proof annotations to pass the proof.

Instead of exhaustively enumerating all possible annotations, \sys exploits the observation that, in practice, the annotations programmers usually add to a failing proof fall into a few commonly used patterns.
\sys uses syntax-guided synthesis (SyGUS)~\cite{Rajeev2013Rastislav} to synthesize these patterns and only generate annotations that help the prover make progress.
By restricting itself to these patterns, \sys keeps the search tractable, even for more complex proofs.

\bubble{We observe the generated lemmas by \sys are independent and can be checked in parallel.}
Once \sys identifies the failing proof obligations and generates candidate proofs for them, it needs to check which candidates actually convince the prover of correctness.
Since these candidate proof annotations can be independently checked by the theorem prover, we extend \sys to evaluate them in parallel by designing a backend that can check these candidate proofs using multiple prover instances. This reduces the execution time of \sys to find the proof of a lemma from multiple hours to a few minutes.

% For a failing proof, \sys generates candidate proof annotations that can be independently checked by the theorem prover\manos{This sentence disrupts the flow}.
% The candidate proof annotations \sys can add to a failing proof are independent of each other.
% To further improve execution time, we design a theorem-prover-as-a-service (\tpas) backend

% that allows \sys to evaluate these candidates in parallel by spawning multiple prover instances.
% This parallel execution significantly reduces the execution time of \sys to find the proof of a lemma (from O(hours) to O(minutes)).

% Thus, \sys can check each of these annotations in parallel by employing different instances of the theorem prover to optimize its execution time.
% This allows \sys to scale and find the correct proof of a lemma in a reasonable amount of time (i.e. reducing from O(hours) to O(minutes)).
% To enable this, we design a Theorem-Prover-As-a-Service (\tpas) backend that can prove these generated lemmas concurrently.
% As a result, \sys allows programmers to trade human time with compute time, utilizing a cluster of nodes to enhance the development of formally verified programs.

% This enables \sys to scale and trade 
% We observe the generated lemmas by \sys are independent due to the unique characteristic of \scopeExpander and can be check in parallel.
% To enable \sys to work at scale, we design a Dafny as a Service backend to

\bubble{We show \sys effectiveness by applying it to variety of distributed protocols safety proof.}
We evaluate the effectiveness of \sys by using it to synthesize safety proofs for a variety of distributed systems protocols, including leader election, Paxos~\cite{Lamport1998}, and Raft~\cite{Ongaro2014Ousterhout}.
For simpler protocols, such as leader election, \sys automatically finds the entire proof in under two minutes, without any human intervention.
For more complex protocols such as Paxos and Raft, \sys helps find the last required proof annotation to convince the theorem prover.
We show that \sys can complete these proofs in one minute to two hours for 86 out of 103 experiments.

% We evaluate \sys on the safety proof of a variety of distributed system protocols\manos{weird phrasing}.
% \sys is able to find the complete proof for protocols such as the leader election protocol\manos{What protocols are these? What will this mean to the reviewers?} without any human intervention in less than two minutes.
% For more complex protocols such as Paxos~\cite{Lamport1998} or Raft~\cite{Ongaro2014Ousterhout}, \sys can find the correct proof in the last steps of protocol verification\manos{Not sure what are the "last steps of protocol verification". If you mean the rest of the sentence, maybe just say that.} where programmer only needs to add one or two more proof annotations.
% Depending on the complexity of the missing proof, it can take from one minute to two hours for \sys to find the missing proof.\manos{We need a better summary of our results.}

In summary, we make the following contributions:
\squishlist
    \item We automate the iterative task of querying the theorem prover and reduce the required human intervention in formal verification.

    \item We introduce a shard-and-explore approach that automatically breaks large proof obligations into smaller ones (proof sharding), allowing us to control the proof exploration of each part independently (proof exploration).

    % \item divide and conquer that allows us to scale and conquer each case separately.
    % \item 
    % \item We identify patterns that programmers typically use as proof annotations. Using these patterns, we employ syntax-guided synthesis (SyGUS) to automatically generate only expressions that match one of the patterns.
    % \item We automate finding the correct proof for a program to convince the theorem prover and thus, reduce human involvement in the proof-debugging process.
    \item We propose \sys, a scalable proof synthesis approach that advances automated proof search using the shard-and-explore workflow.
    % \manos{How is this a separate contribution?}
    % by dividing the proof obligations into smaller pieces where necessary and then finding the proof for each case independently.
    % \item We introduce \textit{Scope Expander} to expand the scope of lemmas where necessary.
    % \item We propose a theorem-prover as a service backend (\tpas) to enable scalability and concurrency for \sys.
    % \item We design a variety of \textit{Lemma Transformers} to generate multiple proof candidates for the same lemma to be checked by the prover.
    % \item We evaluate \sys to show its effectiveness on finding the last necessary proof annotation to finish a proof in a range of one minute to two hours for 81 out of 104 experiments we evaluated \sys on.
    \item We evaluate \sys on 103 proof-completion tasks and show that it finds the last missing proof annotation in 86 cases, with a runtime ranging from one minute to two hours.

    % \item We show \sys can also find the complete proof of leader election protocol from scratch in less than 2 minutes.
\squishend

\section{Background: Writing Proofs Manually}\label{sec:motiv}

\begin{lstlisting}[label=snippet:bookShelfExample,caption=Library example, basicstyle=\small\tt, float=t!,escapechar=$, numbers=left]
predicate Init(s: Library) {
  forall book | book in s :: s[book].Shelf?
}
// definitions omitted due to space
predicate Borrow(
    s:Library, s':Library,
    book:Book, name:string)
predicate Return(
    s:Library, s':Library,
    book:Book, name:string) 
predicate HasAtMostOneBook(
    s: Library, name: string)

predicate Next(s:Library, s':Library) {
  || (exists book, patron :: 
      Borrow(s, s', book, patron))
  || (exists book, patron :: 
      Return(s, s', book, patron))
}
predicate Safety(s:Library) {
  forall name :: HasAtMostOneBook(s, name)
}
lemma SafetyProof()
  ensures forall s | Init(s) :: Safety(s)
  ensures forall s, s' | 
    Safety(s) && Next(s, s') :: Safety(s') 
{ }
\end{lstlisting}

% \armin{should we describe what is safety property here?}
% we prove correctness by proving an invariant. readers familiar with the area that this invariant is usually stronger than the safety property.
% our focus is not what is the inductive invariant. and question is how to prove this invariant is inductive.
To prove safety, programmers must show that the system never reaches an undesirable (unsafe) state.
To achieve this, programmers prove that an invariant holds throughout program execution.
This invariant is usually stronger than the protocol's safety property and is inductive.
Finding this inductive invariant for a protocol is well-studied and prior work~\cite{padon2016ivy, ma2019i4, Yao2022Tao, Yao2021Tao,Zhang2024Hance, Zhang2025Singh} has mostly automated this part of the verification process.

However, after finding the inductive invariant, programmers still need to write proofs.
They prove two main propositions:
% To achieve this, programmers write a proof and prove two main properties:\manos{Are these properties? They serve to prove a safety property, so they can't be}
% \squishlist
(1)~the invariant holds in all valid initial states of the program
and (2)~transitioning from a state \texttt{s} where the invariant holds to the next state \texttt{s'} maintains the invariant.
% \squishend
Once programmers prove these two propositions, they can inductively conclude the protocol maintains this invariant in all reachable states.
% Combining these two properties, programmers can conclude the implemented distributed system is safe in all reachable states.

To illustrate how this works in practice, we use a simple Library protocol shown in Code Snippet~\ref{snippet:bookShelfExample}.
% The safety property we want to prove is that each patron can hold at most one book at any time.
This protocol models the operations of borrowing and returning books in a Library as a state machine.
Initial states of this protocol are defined by the \texttt{Init} predicate on line 1.
This predicate specifies that all books should start on the shelf initially.
Next, the \texttt{Borrow} and \texttt{Return} state machine transitions are defined in lines 5 and 8, respectively, as predicates using the current state (\texttt{s}) and next state (\texttt{s'}).
The \texttt{Next} predicate at line 14 defines the valid transitions that this state machine can take in each step (\texttt{Borrow} and \texttt{Return}).

% An example of a lemma proving these two properties for proving safety is depicted in line 23 of Code Snippet~\ref{snippet:bookShelfExample} for a bookshelf example.
% Programmer first defines protocol behavior using state machines.
% The initial states are defined in predicate \texttt{Init} in the first line of this code snippet, which states all books in the Library are on the shelf initially.
% The next step is defined in line 14 as the disjunct of two possible transitions, a patron borrowing from or returning a book to the Library.

After defining the protocol, programmers specify the safety property and prove that all reachable states maintain this property.
% to specify what states are safe for the protocol to reach\manos{weird phrasing}.
The safety property in the Library example is defined in line 20, declaring that each patron can have at most one book at any point in the protocol's execution.
Lastly, the programmer writes the \texttt{SafetyProof} lemma on line 23 to inductively prove all reachable states in this program are safe.
% This lemma does not automatically verified by the theorem prover as it requires

% \manos{You need to make it clear that this is not end. When you say "to prove", they might understand that this is how you prove it.}

% \subsection{Proof Development Cycle}
Ideally, after writing the \texttt{SafetyProof} lemma, the programmer would send this program to the theorem prover and it would automatically prove the safety lemma.
In practice, however, writing a lemma like \texttt{SafetyProof} is only the beginning of the proof development cycle.
Theorem provers often fail to prove the correctness of even simple protocols such as the library example in Code Snippet~\ref{snippet:bookShelfExample}.
To convince the theorem prover of the correctness of this lemma, the programmer needs to write extra proof annotations.
% \manos{I think you can make a stronger point here. This lemma is a good example of how even not-so-complex lemmas don't automatically verify.}\armin{fixed}

After the theorem prover fails to prove this lemma, its only guidance to the developer is that it could not prove the post-condition on line 25.
% Also, theorem provers only accept a single proof to check at a time\manos{This feels forced and out-of-place}.
% \manos{How is this a result of the previous sentence?}\armin{that the guidance from the theorem prover is small?}
Without more direct guidance from the theorem prover, it is up to the developer to discover the root cause of the proof failure.
The programmer must \textit{manually} determine what proof annotation to add next and query the theorem prover again.
% This requires expertise to determine what proof annotation to try next and is a barrier in adopting formal verification in practice.
% \manos{What guidance do they provide? We can't be vague at this point}\armin{fixed}

% This back and forth\manos{This is a bit weird, since you've mentioned the back and forth a few times now. Since we have a concrete example, I was hoping you would have shown the actual back and forth before making this point. Not sure what is the point of this paragraph.} between the theorem prover and programmer is common not only in the Library example, but across most distributed systems.
% This makes formal verification time-consuming and is a barrier to adopt formal verification in development cycle.\manos{It's OK to repeat important points, but this is now the third time (in two pages) we've made this point---and without any evidence to back it up.} 

% To show why this task is challenging, we next walk through how a programmer completes the safety proof after the initial attempt fails, and what type of proof annotations they add to make progress and convince the theorem prover.

\subsection{Proof Annotations}
Based on the theorem prover's response after the first attempt to verify the \texttt{SafetyProof} lemma,
% \manos{you haven't actually said that this fails, yet}\armin{fixed}
the programmer concludes that proving the induction step of the safety proof needs further proof annotations.
To make progress, programmers must manually add annotations to guide the theorem prover to prove the induction step.
In the rest of this section, we go through a step-by-step approach of what proof annotations programmers usually add to this program for the proof to pass.

\begin{table}[]
\centering
\caption{Selected triggers by the prover in the Library example.
To instantiate a quantifier, programmer needs to invoke one of the trigger sets assigned to this quantifier. For instance, to instantiate the quantifier in line 25, programmer can either invoke \texttt{Next(s, s')} or both \texttt{Safety(s)} and \texttt{Safety(s')}.
% \manos{What do two entries in the same row mean?}}\armin{multiple triggers that each will instantiate the quantifier. line 25 first trigger is a set, so programmer need to specify both Safety(s) and Safety(s') to instantiate.
}
\begin{tabular}{|c|cc|}
\hline
2  & \multicolumn{1}{c|}{\texttt{s{[}book{]}}}               & \texttt{book in s}   \\ \hline
15 & \multicolumn{2}{c|}{\texttt{Borrow(s, s', book, patron)}}             \\ \hline
17 & \multicolumn{2}{c|}{\texttt{Return(s, s', book, patron)}}             \\ \hline
21 & \multicolumn{2}{c|}{\texttt{HasAtMostOneBook(s, name)}}               \\ \hline
24 & \multicolumn{1}{c|}{\texttt{Safety(s)}}                 & \texttt{Init(s)}     \\ \hline
25 & \multicolumn{1}{c|}{\texttt{\{Safety(s), Safety(s')\}}} & \texttt{Next(s, s')} \\ \hline
\end{tabular}
\label{tabular:triggers}
\end{table}

\subsubsection{Introducing new variables}
A common first step in verifying this lemma is to introduce new variables to reason about.
In the Library example, the programmer can use the following \texttt{forall} statement to introduce variables \texttt{s} and \texttt{s'} to write proofs on the state transition in \texttt{Next}.
\begin{lstlisting}[label=snippet:expandScope1, basicstyle=\small\tt,escapechar=$]
forall s, s' | Safety(s) && Next(s, s')
ensures Safety(s') { }
\end{lstlisting}
% This is a common strategy in writing proofs to introduce new variables to the scope and use them to write more proofs inside the \texttt{forall} statement.

Simply adding this proof annotation, however, is not sufficient to convince the theorem prover of the correctness of the safety proof.
At this stage, the programmer has two options to make progress.
% explore finding the correct proof for this lemma.
First, they can replace \texttt{Next(s, s')} with the definition of the \texttt{Next} predicate, which is a disjunct of two possible state transitions.
This helps programmers identify whether the theorem prover can automatically prove one of these state transitions correct.
In that case,
% \manos{in what case? you mentioned both}\armin{in case one of the transitions is automatically proven correct}
programmers can focus on only the cases that require further proof annotations.

Alternatively, the programmer can replace \texttt{Safety(s')} in the post-condition (i.e., \texttt{ensures}) of the \texttt{forall} statement with the definition of the \texttt{Safety} predicate.
\begin{lstlisting}[label=snippet:expandScope2, basicstyle=\small\tt,escapechar=$]
  forall name
  ensures HasAtMostOneBook(s', name)
  { }
\end{lstlisting}
This replacement lets the programmer introduce another variable, \texttt{name}, into the program scope.
Introducing this variable turns out to be crucial for finishing this proof as we explain later in this section.
However, programmers cannot know in advance which of these two options is better to pursue first. 
Hence, they need to explore both approaches and engage in an iterative exchange with the prover until the proof is complete.

Of course, the more complicated the proof, the more such ``forks'' there are for the developer to explore, which greatly increases the required manual effort in traditional proof-debugging.

\subsubsection{Triggers}\label{sec:trigger}
Another common source of manual effort arises from quantifiers.
Even after expanding definitions and exploring different proof cases, programmers often find that the prover still cannot complete the proof.
To understand why, lets take a closer look at how theorem provers handle quantifiers such as $\forall x :: P(x) \Rightarrow Q(x)$. Assume $x$ is an integer.
Provers do not blindly instantiate $x$ with every possible value, as that would generate an infinite number of facts of the form $P(x) \Rightarrow Q(x)$.
Instead, they rely on \emph{triggers}: a set of expressions associated with each quantifier that determines when that quantifier should be instantiated.
Theorem provers only instantiate a quantifier whenever all expressions in the trigger set of this quantifier are mentioned in the program scope.
For instance, the trigger for the above quantifier can be either $P(x)$ or $Q(x)$ or both.
% Lets assume the selected trigger by the theorem prover for this quantifier is $P(x)$.
% If programmer wants to instantiate this quantifier with a particular value $A$, they need to specifically mention $P(A)$ in their proof to guide the theorem prover on instantiating this quantifier.

%\manos{I think you need an introductory sentence to maintain the flow. Something like "another source of manual effort for developers is in finding the required \em{triggers}. Triggers are a way to overcome the blah blah problem..." Or maybe don't mention triggers but definitely mention this is another source of manual effort you are about to introduce.}
% Theorem provers do not instantiate quantifiers with all possible values for bounded variables to avoid search space explosion.
%\manos{This statement is too dense to understand for anyone that doesn't already know what triggers (or even quantifiers) are. You need an example.}
% Instead, they rely on \textit{triggers}: a set of expressions associated with each quantifier that determines when that quantifier should be instantiated.
% Each quantifier is assigned an expression set, called trigger.
% Theorem provers only instantiate a quantifier whenever all expressions in the trigger set of this quantifier are mentioned in the program scope.

The trigger set for each quantifier is usually assigned automatically by the prover, although the programmer may also select these triggers manually.
Table~\ref{tabular:triggers} shows the selected trigger sets by the prover for all quantifiers in the Library example.
Programmers need to manually invoke a quantifier's triggers to guide the theorem prover in instantiating the quantifier with particular bound variables.

% The quantifier in line 25 is only instantiated whenever the predicate \texttt{Safety} is mentioned for any two variables \texttt{s} and \texttt{s'}.
For instance, the trigger for the quantifier in line 21 of Code Snippet~\ref{snippet:bookShelfExample} is \texttt{HasAtMostOneBook(s, name)}.
For this proof to pass, programmers must instantiate this quantifier using variables \texttt{s} and \texttt{name} as inputs of the \texttt{HasAtMostOneBook} predicate.
Otherwise, the theorem prover will not instantiate this quantifier automatically and will fail to prove this lemma.
Upon invoking this trigger, the theorem prover can automatically conclude that the safety lemma is correct.
The complete proof for the \texttt{SafetyProof} lemma is shown in Code Snippet~\ref{snippet:bookShelfProof}.

\begin{lstlisting}[label=snippet:bookShelfProof,caption=Safety proof for the Library example, basicstyle=\small\tt,
float=t!,escapechar=$, numbers=left]
lemma SafetyProof()
  ensures forall s | Init(s) :: Safety(s)
  ensures forall s, s' | 
    Safety(s) && Next(s, s') :: Safety(s') 
{
  forall s, s' |
    Safety(s) && Next(s, s')
  ensures Safety(s') {
    forall name
    ensures HasAtMostOneBook(s', name) {
      var trigger :=
        HasAtMostOneBook(s, name);
    }
  }
}
\end{lstlisting}

Identifying the correct trigger to invoke is a difficult and mundane task. 
A large portion of time spent for writing proofs by programmers is on finding the correct invocation of a trigger~\cite{Becker2019, Leino2016Pit-Claudel}.
Note that in most cases, finding the necessary trigger invocations cannot be done in isolation.
The programmer may not be able to invoke the necessary triggers for the proof to pass without introducing new variables into the program scope.
% may require the new variables added by scope expansion, as shown in the bookshelf example.
Moreover, only relying on adding one type of proof annotation alone does not yield a correct proof.
Instead, programmers must combine several annotations in the right order, which requires expertise and repeated interaction with the theorem prover.
% A combination of different proof annotation types is usually required to complete the proof of a lemma.
% The goal of this work is to reduce the human involvement in the back and forth cycle of adding proof annotations and querying the theorem prover.

% Automatically adding proof annotations and querying theorem prover while removing human from the cycle of proof development can significantly help adopting formal verification in practice.

% Automatically adding proof annotations to failing lemmas and querying the theorem prover helps adopting formal verification in software development.

\section{Narrow \& Explore}\label{sec:design}
Our goal in this work is to reduce the amount of human intervention required in the mundane task of iterating on the proof to find the correct set of proof annotations needed to prove correctness.
The proposed design should be able to automatically generate proof annotations that programmers write manually when the theorem prover fails to prove correctness.
Since the number of possible annotations that can be added to a lemma is unbounded, a naive approach of simply adding proof annotations to a failing proof one by one is not scalable.
% In addition, distributed systems exhibit different transitions during their execution where each may need different proof annotations to prove correctness.
% Attempting to find the proof of correctness in one iteration without going back and forth to the theorem prover is not possible.

A key observation is that programmers typically do not prove lemmas in a single attempt.
Instead, they query the theorem prover repeatedly, using its feedback to decide which annotation to try next.
% Our design embraces a similar workflow and complements the theorem prover.
% We rely on the prover to check if a proof is correct or not, and focus on generating the proof annotations that the programmer typically adds in each iteration of querying the theorem prover.
% The goal is not to reinvent the wheel and mimic what theorem prover does underneath.
% Our goal is to .
% Rather than reinventing the wheel and attempting 
% Just like how programmers attempt to find proof of correctness by manually querying the theorem prover back and forth, our proposed design also should leverage the automation provided by the theorem prover.
% Instead, our goal is to complement the theorem prover.
% Our focus is to find the proof in cases where the theorem prover fails to automatically prove correctness.
The annotations that programmers typically add in each iteration fall into two main categories: (1)~narrowing the scope of the proof and (2)~exploring the proof for each scope.
Our automated design for generating annotations follows a similar approach: it should be able to add both types of annotations automatically wherever possible and iterate on the proof.
As discussed in Section~\ref{sec:motiv}, relying on only one type of proof annotation is rarely enough to find the proof of correctness.
Therefore, our design uses two key components to generate each class of proof annotations: \textbf{Narrow} and \textbf{Explore}.

% To achieve this goal, we need to add proof annotations that theorem provers do not explore on their own.
% We first identify the pattern of proof annotations that programmers usually add to convince the theorem prover.
% Using these patterns, we generate possible proof annotations to add to the proof using controlled exploration.
% We then use theorem prover to check if it can prove correctness using the augmented proof (i.e., original proof in addition to the generated proof candidates).

% As a result, we use two key components in our design to (1) divide the proof obligations wherever possible and (2) conquer each obligation separately.
% This strategy helps finding the correct proof of correctness in multiple ways.

\textbf{Narrow.} 
When possible, we shrink the proof obligations of a lemma into smaller, independent cases.
This helps break down large proofs into smaller chunks that require fewer and simpler proof annotations to pass.
% In addition, it allows us to explore the sufficient proof annotation for each case individually.
% Also, this isolates different parts of the program that need distinct proof annotations, and let the theorem prover only focus on one case at a time, reducing the chance of search space explosion.

% Lastly, the divide phase helps us conquer the correct proof of correctness for each case more effectively.

% The divide phase helps us focus on only a portion of the proof, and check possible proof annotations that can be added to this part of the proof individually.

\textbf{Explore.}
% To find the proof for each proof obligation, we need to avoid generating proof annotations that are futile to the proof.
After narrowing the scope of the proof, for each shard, we begin searching for proof annotations that help the theorem prover make progress while avoiding annotations that are futile to the proof.
For instance, in the library example, the expression \texttt{v[book].Shelf?} is not a trigger for any of the quantifiers in the code, as shown in Table~\ref{tabular:triggers}.
As a result, generating expressions of the format \texttt{v[book].Shelf?} is pointless because they will not change the theorem prover's behavior, whereas expressions of the format \texttt{v[book]} may change the theorem prover's behavior.
By focusing proof exploration only on patterns that affect the theorem prover's behavior, we avoid wasting effort on annotations that have no effect.

Together, the narrow and explore principles enable controlled exploration of the proof search space.
We break a large verification task into smaller pieces, and then explore the proof search space for each chunk using controlled search-space exploration.

\section{ProofSaX}

Using the shard-and-explore principles explained in \S\ref{sec:design}, we design \sys to complete the proof of a lemma when theorem provers require manual input from developers.
% We use a \textit{Divide} and \textit{Conquer} approach to find proofs of failing lemmas.
At the core of \sys are a set of \textit{Lemma Transformers}, each of which takes a failing lemma and adds one or more proof annotations to this lemma.
% for the theorem prover to check.
These transformers work together: some focus on sharding the lemma into smaller chunks (proof sharding), while others focus on adding proof annotations that cause the prover to investigate the proof search space further (proof exploration).
% that given a failing lemma, they generate a new candidate lemma to be checked by the theorem prover.
% Lemma Transformers generate lemmas such that if a  If a child lemma is passed the proof of correctness by the theorem prover, 
% The combination of these lemma transformers work together to find the correct proof of a lemma.
% using the divide and conquer approach.

\begin{figure}[t!]
\centering
\includegraphics[width=\columnwidth]{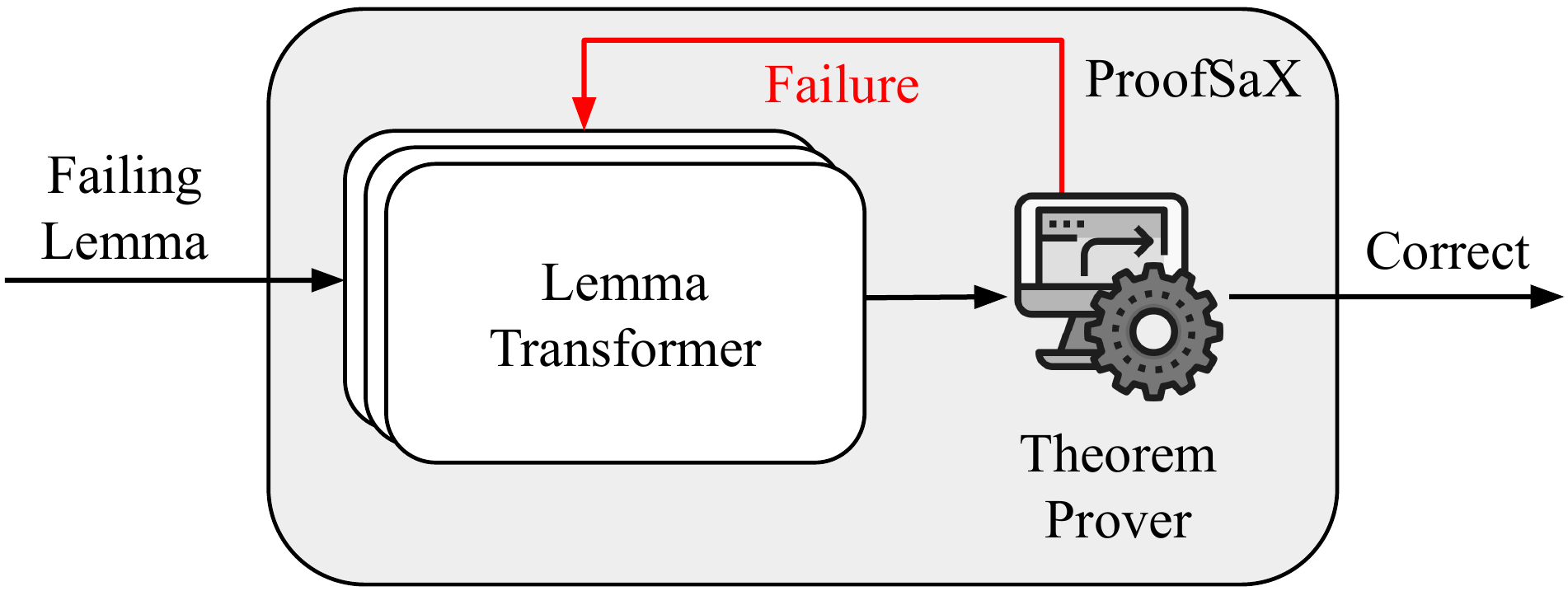}
\caption{
High-level overview of \sys: For a failing lemma, \sys generates candidate proofs using lemma transformers. It then checks the candidate proofs using the underlying theorem prover. If it successfully finds a passing proof, it returns the correct proof to the programmer. Otherwise, it recursively refines the failing lemma using lemma transformers designed according to the shard-and-explore principle.
}
\Description{
High-level overview of \sys: For a failing lemma, \sys generates candidate proofs using lemma transformers. It then checks the candidate proofs using the underlying theorem prover. If it successfully finds a passing proof, it returns the correct proof to the programmer. Otherwise, it recursively refines the failing lemma using lemma transformers designed according to the shard-and-explore principle.
}
\label{fig:sys}
\end{figure}

Figure~\ref{fig:sys} shows an overview of \sys.
Given a failing lemma, lemma transformers generate new candidate proofs for this lemma.
% to be checked by the theorem prover.
The candidate proofs are then sent to the theorem prover to check their correctness.
If a correct proof is found, \sys returns the proof to the developer.
Otherwise, these candidate lemmas are used to generate new proofs using the lemma transformers and shard/explore even further.

Table~\ref{tabular:transformers} shows the list of lemma transformers we designed in \sys.
The first three lemma transformers 
aim to explore the proof search space by adding triggers, invoking lemmas, or revealing opaque definitions in the proof.
% by adding annotations that may change the theorem prover behavior and ideally pass the check by the theorem prover.
% On the other hand, the 
The last two lemma transformers shard the proof obligations and narrow the scope of the proof by expanding either pre-/post-conditions of lemmas or quantifiers.
% \sys uses a combination of these transformers to make shard and explore the proof search space.
% \sys can find the proof of lemmas by using a combination of these transformers.

Next, we first describe how \sys controls the exploration space of proof annotations to avoid search-space explosion, and then explain how each lemma transformer contributes to finding the proof of a failing lemma.
% then explain each of the lemma transformers in more detail.

\begin{table}[]
\centering
\caption{List of Lemma Transformers}
\begin{tabular}{|c|c|}
\hline
Name                       & Type    \\ \hline
Trigger Synthesizer             & proof exploration \\ \hline
Lemma Invocation                & proof exploration \\ \hline
Reveal Opaque Definitions       & proof exploration \\ \hline
Expanding Pre-/Post-conditions  & proof sharding  \\ \hline
Quantifier Elimination          & proof sharding  \\ \hline
\end{tabular}
\label{tabular:transformers}
\end{table}

\subsection{Controlled Expression Synthesizer}
% \sys controls the number of possible expressions it generates at each step in two ways.
Since the number of possible expressions that can be added to a lemma as proof annotations is unbounded, \sys controls the synthesis of expressions based on their complexity.
We define the complexity of an expression based on the number of its operands and variables.
For example, consider the expression \texttt{s[book].Shelf?}.
The variables \texttt{s} and \texttt{book} are unit expressions of complexity one.
Since \texttt{s[book]} is the combination of the two, the complexity for \texttt{s[book]} is the sum of each of its operands' complexity plus the indexing operator, yielding a complexity of three.
Moreover, accessing an element in a datatype adds one to the complexity, which makes \texttt{s[book].Shelf?} an expression of complexity four.

\begin{figure}[t!]
    % \centering
    % $c(x) = complexity\ of\ x$ \\
    \begin{align}
    & c(OP\ a) = c(a) + 1 \tag{C-Unary}\\
    & c(a\ OP\ b) = c(a) + c(b) + 1 \tag{C-Binary}\\
    & c(s[i]) = c(s) + c(i) + 1 \tag{C-Collection}\\
    & c(a.b) = c(a) + 1 \tag{C-Datatype}\\
    & c(foo(a, b)) = c(a) + c(b) + 1 \tag{C-Function}
    \end{align}
    \caption{Expression complexity calculation formulas}
    \Description{Expression complexity calculation formulas}
    \label{fig:exprComplex}
\end{figure}

More formally, the complexity of expression $x$, namely $c(x)$, is calculated using the formulas depicted in Figure~\ref{fig:exprComplex}.
Unary operations such as negation, or cardinality of a set, increase the complexity by one (C-Unary).
The complexity of a binary operation expression is the sum of each operand plus one (C-Binary).
Similarly, accessing element $i$ from a collection $s$ has the combined complexity of $s$ and $i$ plus one for the indexing operand (C-Collection).
Accessing an element of a datatype increases the complexity by one (C-Datatype).
Function invocation complexity equals the sum of its arguments plus one (C-Function).

% of the format $E$ with maximum expression complexity of $c(E)$
To implement the controlled synthesis of expressions, we design \exprGenerator.
Initially, \exprGenerator gathers all unit variables in the scope of the lemma at the point of failure, including lemma arguments and locally declared variables.
\exprGenerator determines the syntactic type of each variable, and maintains a mapping $M$ from their types to the unit variables.
After finding any new expression\manos{What does this mean? How does this happen?}, \exprGenerator inserts an element into this mapping with the syntactic type of the generated expression as the key and the new expression itself as the value.

Next, there are two possible cases\manos{This is too procedural. I don't know where this is going.} when synthesizing an expression of the syntactic format $F$ with maximum expression complexity $c(F)$ using these unit variables.\manos{What is format F and where does a maximum expression complexity come from?}

\manos{Don't write these like an algorithm. Explain the insight behind this.}
First, $F$ is a syntactic expression format that cannot be broken down to smaller expression formats.
% m unit expression itself and not possible to be broken down to smaller expressions.
In this case, \exprGenerator looks up the syntactic type of $F$ in mapping $M$ and returns all expressions in the mapping with this syntactic format.

Second, $F$ can be a combination of some smaller syntactic type $f_1, ..., f_n$ with operation $O$ on them (e.g. binary operator, function call).
To synthesize an expression of format $F$, \exprGenerator recursively synthesizes expressions of the format $f_1, ..., f_n$ with maximum complexity $c(F) - n - 1$.\manos{Why?}
Lastly, \exprGenerator computes the cross-product of all options whose overall complexity is less than $c(F)$ to synthesize expressions of the format $F$.\manos{Again, why?}

% \sys implements this controlled expression synthesis through \exprGenerator in three steps.

% Second, it maintains a mapping from the type of each expression to all generated expressions of that type so far.
% Each newly generated expression is added to this mapping.
% Given the syntax of the expression that we want to generate and its maximum complexity, \exprGenerator recursively generates all sub-expressions in that syntax until it reaches the unit variables. 
% Third, because the syntax of generating expressions is usually shared between all lemma transformers, \exprGenerator caches generated expressions by their type and complexity.
% The result is a set of expressions that transformers can reuse while keeping the exploration space bounded and incremental.

% To implement the controlled synthesis of expressions, we design \exprGenerator.
% First, \exprGenerator generates the list of all unit variables available in the scope of the lemma, including lemma inputs and declared variables until the point of failure.
% Second, \exprGenerator maintains a mapping from the type of expressions to all expressions of that type.
% Each newly generated expression is added to this mapping.
% Lastly, given the syntax of the expression we want to generate and its maximum complexity, \exprGenerator recursively generates all sub-expressions in that syntax until it reaches the unit variables.
% \exprGenerator caches the generated expressions for each type and expression complexity as the syntax of generating expressions is usually shared between all lemma transformers.

Next, we discuss how each proof-exploration lemma transformer uses \exprGenerator to synthesize proofs.

\subsection{Trigger Synthesizer Transformer}
% \manos{You've explained this before. Mention this here.}
% to avoid search space explosion.
% Instead, they rely on triggers.
% Each quantifier in the code is assigned a trigger, either automatically by the prover or manually by the programmer.
% In order to instantiate quantifiers, programmers need to manually invoke the triggers of a quantifier.
% Programmers need to determine what expression to add to their proof that triggers the necessary instantiation of a quantifier to convince the theorem prover on the correctness of their program.\manos{Do you need to re-explain all this?}
As we described in \S\ref{sec:trigger}, theorem provers use triggers to handle quantifiers.
To automate the generation of trigger invocations, we first explain how an expression is internally matched to a trigger in the theorem prover.
Theorem provers use the E-matching~\cite{Detlefs2005Nelson} algorithm to check if an expression in the program matches the trigger of a quantifier or not.
An expression is matched to a quantifier if and only if they both follow the same pattern.\manos{Wait, if that's the criterion, then why did you mention E-matching?}\armin{They use E-matching algorithm to determine if two expressions match or not}\manos{Also, what does "exact same" mean here?}\armin{fixed}
For instance, the trigger \texttt{s[book]} in Table~\ref{tabular:triggers} may only match expressions of the format $map[string]$ where an element of a map is accessed.

We leverage the E-matching algorithm to generate only expressions that can match the trigger of at least one quantifier.
Starting from the syntax of all triggers defined in the program, \sys utilizes syntax-guided synthesis (SyGUS)~\cite{Rajeev2013Rastislav} to generate the possible triggers.
First, \sys uses \exprGenerator to generate all possible expressions that syntactically match a particular trigger with minimum expression complexity.
If it turns out that the minimum expression complexity does not suffice in finding the correct proof, \sys then increases the target expression complexity by one until it finds the correct proof or reaches maximum expression complexity.
\manos{What does "step-by-step" mean? Also, do you increase the expression complexity or the *target* complexity?}\armin{fixed}
% Next, we will generate a new candidate lemma and query the theorem prover.

\manos{The flow seems completely disrupted here. What is the new topic? Maybe add a bf header?}
\textbf{How to invoke triggers?}
Adding trigger invocations to a failing lemma is not trivial, and we need to answer a key question for this task.
Should we generate one lemma per trigger invocation?
\manos{I thought we wouldn't use the term "trigger instantiation"}\armin{is trigger invocation good?}
Or just one lemma that includes all invocations? Or something in between? 
% Second, how should \sys handle lemmas where multiple trigger invocations are required at the same time in order to complete the proof?
\manos{I'm not sure this question is easy to follow. Also, it's not clear this is a separate question. Consider not structuring this as a question, but describing the naive/strawman solution as generating one lemma per trigger and then saying why that's bad. Then you can say how we can do better.}\armin{removed. explained in next paragraph}

The number of trigger invocations generated by \exprGenerator grows exponentially with increasing complexity.
Hence, generating a lemma per trigger invocation is not scalable and requires sending a large number of queries to the theorem prover.
In addition, this naive solution does not cover cases where more than one trigger invocation is required to complete the proof.
At the other end of the spectrum, adding all trigger invocations to a single lemma and only querying the theorem prover once leads to search-space explosion.
This approach \manos{Why do you say "though"? This doesn't seem to contradict the previous sentence. It agrees with it.}\armin{removed} directly contradicts the design principles of theorem provers and defeats the purpose of triggers in theorem provers.

\begin{figure}[t!]
\centering
\includegraphics[width=\columnwidth]{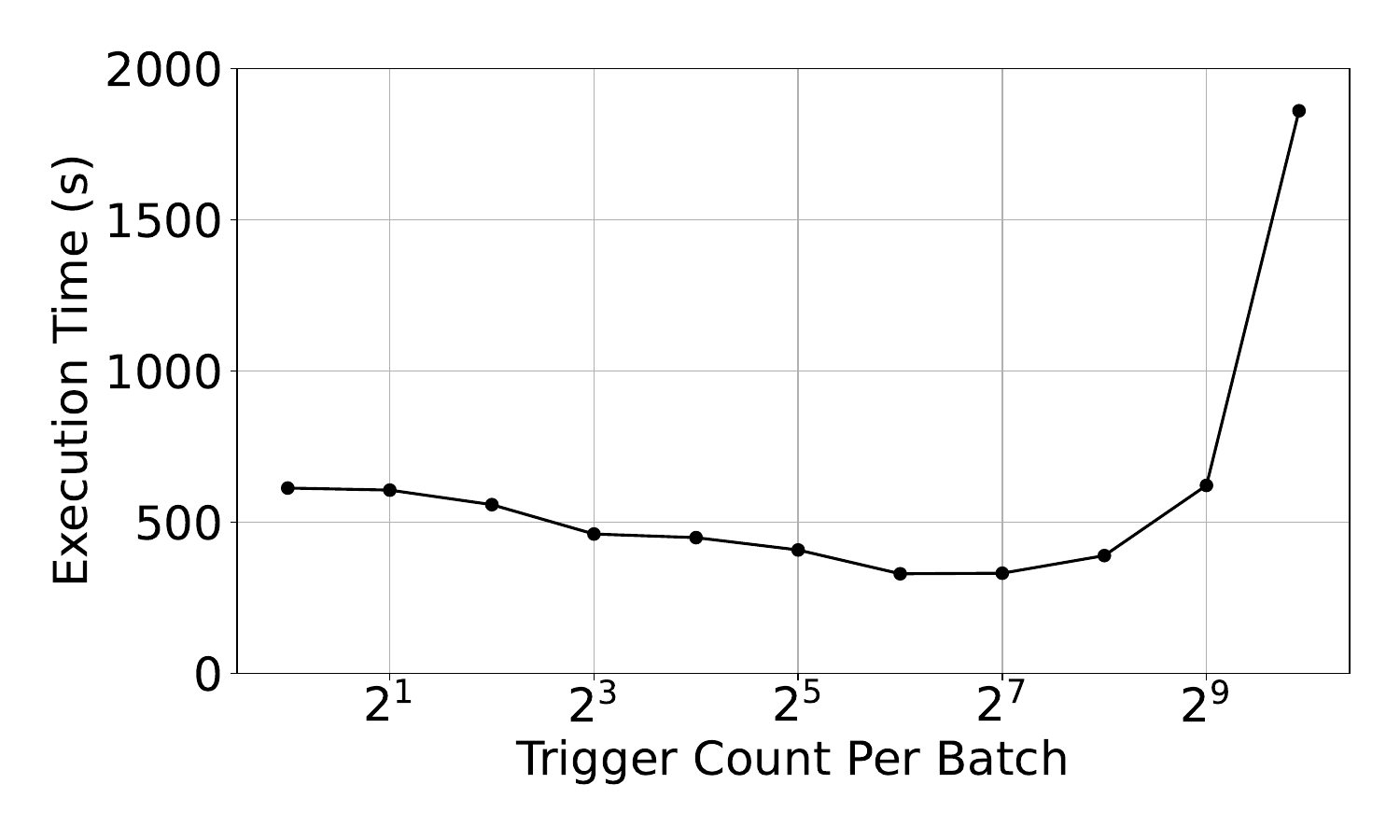}
\caption{Overall execution time of checking 1024 generated trigger invocations when added to a lemma in batches of size 1 to 1024.}
\Description{Overall execution time of checking 1024 generated trigger invocations when added to a lemma in batches of size 1 to 1024.}
\label{fig:trigger_analysis}
\end{figure}

\manos{Rephrase this sentence. Something like "To resolve this question in practice, we measured...". Btw, I think past tense is appropriate here, since this analysis was performed before the stuff we present in the paper, since it causally precedes it. Also, this works best if you have already presented the above as a question.}
To resolve the above question, we analyzed how many trigger invocations per query yield the fastest overall execution time when checking 1024 generated trigger invocations for a lemma.
Figure~\ref{fig:trigger_analysis} shows the overall execution time when checking all triggers in batches of size $k$, where $k$ ranges from 1 to 1024.
As shown in this figure, adding 64 triggers per lemma yields the lowest overall execution time. Of course, a batch size of 64 is not a silver bullet; merely an empirically derived happy medium. 

\manos{This has again become a linear story whose ending I don't know. Also, why is this "Next"? After what?} 
\textbf{Multiple Trigger Invocations:}
We need to check cases\manos{What does that mean?} where multiple trigger invocations are required to prove a lemma.
Based on empirical results\manos{What results? Do you show them?}, we have not encountered any lemma where three or more trigger invocations are required to convince the theorem prover of correctness \textit{after} proof sharding.
As a result, we only test tuples of size two.\manos{I think this discussion should have been earlier.}

% The middle-ground approach of adding multiple trigger instantiations at a time for each query to the theorem prover yields to fastest overall execution time.
% However, that does not cover cases where a lemma may need multiple trigger instantiations to pass.
% As a result, this approach requires generating a large number of lemmas to cover any combination of generated trigger instantiations.
% On the other hand, adding all generated triggers to a single lemma will cover all combinations of trigger instantiations but this leads to a large lemma for the theorem prover to process and explodes the search space.
% We, therefore, decided to add triggers in batches and make sure each tuple of $l$ triggers appear at least in one batch.

At the high level, this is a covering problem~\cite{schonheim1964coverings}.
Given $n$ triggers, we want to generate the smallest number of batches of 64 triggers such that each tuple of two triggers appears at least once in one of the batches.
This is defined as $C(n, 64, 2)$.
WLOG, we assume $n$ is a power of two $(2^k)$ and pad up to the next power of two.
% in each batch of 64 in the covering, we leave the extra instantiations that do not exist as blank.
\manos{I think this can be phrased better. Do you mean you "pad up to the next power of two" or "up to 64"?}\armin{fixed}
% WLOG, we assume $n$ is a power of 2 ($2^m$). If the number of generated triggers by \exprGenerator is $t$, we consider first power of 2 greater than or equal to $t$ to generate the trigger batches.
To generate the $C(2^k, 64, 2)$ covering that satisfies the above property, we recursively generate the $C(2^{k-1}, 32, 2)$ covering.
Using this covering, we generate the $C(2^k, 64, 2)$ covering as follows: (1) two sets of batches using $C(2^{k-1}, 32, 2)$ for the first and second halves of $2^k$ expressions.\manos{not sure what that means}
This covers all tuples of size two within each half.
(2) The cross-product of every 32 triggers in the first and second batch to generate batches of 64\manos{also not sure what this means}.
We generate the $C(2^k, 64, 2)$ covering for a variety of $k$ once (offline). \sys then uses these coverings to generate all lemmas that cover every tuple of size two.

% To control the exploration search space, we only add limited number of triggers per lemma to check by the theorem prover, k. We also generate a covering of triggers such that all tuples of size m are appeared at least once in one batch.
% Assume we have n triggers and want 

% \sys does not generate a lemma for each of these expressions and instead, it adds all the generated expressions to the point of failure in the proof at the same time.
% Although this will increase the necessary time for the theorem prover to process this new lemma, this allows \sys to explore cases where multiple trigger instantiations are necessary for a proof to pass at the same time.

% If \sys is successful in finding the proof of a lemma after adding all trigger instantiations, we provide the proof with all instantiations to the programmer.
% Programmer can then use binary search to determine which of the generated trigger instantiations is in fact necessary and which are futile to minimize the size of the final proof of their program.

% If current expression complexity turns out to be insufficient to complete the proof, \sys increases the expression complexity one by one to try more complex expressions.
% This controlled exploration of triggers helps \sys navigate the proof search space incrementally.

% Witness Synthesizer Transformer works similar to the trigger synthesizer transformer where they both use the \exprGenerator to generate expressions based on the syntax of triggers.
% Witness Synthesizer generated expressions based on the existential quantifiers' trigger syntax while trigger synthesizer

\subsection{Lemma Invocation Transformer}
\manos{Rephrase this sentence}Just as programmers use helper functions to reuse already written functionality in other programming languages, they use helper lemmas in formal verification.
\sys identifies the lemmas that can be invoked at the point of failure in a lemma without introducing loops in the function call graph.
\sys then generates invocations of these helper lemmas by relying on \exprGenerator to generate arguments of the lemma invocation.
\sys creates a new lemma for each of these lemma invocations and passes it to the theorem prover to check its correctness.
If any of the new lemmas generated by this transformer passes, we skip running the remaining lemmas and pass the correct proof to the programmer.

\subsection{Reveal Transformer}
\manos{In the absence of more background on this in Section 2, the following description is not enough for some to understand opaque/reveal.}
Theorem provers, by default, can expand all definitions in the program to explore the proof search space.
However, in some cases, this leads to search-space explosion as theorem provers may frequently expand a definition that is unrelated to the proof or not useful for proving correctness.
As a result, hiding such definitions is a common practice used by developers to avoid this issue.
\manos{What does this mean? The automation in the theorem prover is presumably beyond the developer's control.}\manos{Nit: we have been using both "programmer" and "developer". We should probably use only one of them.}
Developers mark these definitions as opaque in the code and reveal them only where necessary.
Programmers may try revealing these definitions in the program as part of the proof-debugging loop to convince the theorem prover of the correctness of their program.
The Reveal Transformer does exactly that automatically.
This transformer relieves the programmer from the burden of checking this manually and is one of the most effective transformers in \sys, as we will show in \S\ref{sec:eval_single_hole}.

Next, we will explain the lemma transformers that shard the proof into smaller chunks.
% These transformers usually do not change the theorem prover behavior as they only expand definitions, and pre-/post-conditions.
% However, they are an essential part of \sys to complement the conqueror lemma transformers.

\subsection{Pre/Post-condition Sharder}
A common technique used by developers to debug their proofs is to break down the proof into smaller pieces to determine which part of the proof needs extra proof annotations to convince the theorem prover.
This can be done by recursively breaking down the pre-conditions or post-conditions of a lemma.
\sys uses this technique\manos{this makes it sound like there is nothing novel. We do this a lot more methodically and aggressively than the developer, though.} to automatically break down the proof into smaller pieces, which lets the exploration lemma transformers focus on finding the missing proof annotations for each shard separately.
\manos{What does "directly" mean here?}\armin{fixed}

\begin{lstlisting}[label=snippet:preCondition,caption=Breaking a lemma into multiple lemmas based on disjuncts in pre-condition, basicstyle=\small\tt,
float=t!,escapechar=$, numbers=left]
lemma NextPreservesSafety(
    s:Library, s':Library)
  requires Next(s, s')
  requires Safety(s)
  ensures Safety(s')
{
  if (exists book, patron :: 
    Borrow(s, s', book, patron)) {
    // sufficient proof annotations
    // to prove correctness
  } else if (exists book, patron :: 
    Return(s, s', book, patron)) {
    // no proof annotation
  }
}
\end{lstlisting}

At a high level, \sys shards disjuncts/conjuncts in pre-/post-conditions of lemmas, respectively.\manos{This was too short for me to understand.}

\textbf{Sharding Pre-conditions:}
\sys breaks down the pre-conditions of a lemma whenever there is a disjunct in a pre-condition, and generates a new lemma for each case.
For instance, the pre-condition \texttt{Next(s, s')} in Code Snippet~\ref{snippet:preCondition} consists of two disjuncts based on two protocol transitions (Borrow and Return) in the protocol.
\sys generates two candidate lemmas for this example by replacing the original pre-condition \texttt{Next(s, s')} with the two existential quantifiers in the \texttt{Next}\manos{Should this be tt?}\armin{fixed} predicate definition, shown in line 14 of Code Snippet~\ref{snippet:bookShelfExample}.

\sys keeps the proof of the original (failing) lemma unchanged in the new lemmas.
Reusing already written proofs by the developer is important to avoid doing unnecessary work if the programmer has already proven correctness for one of the protocol transitions.
For example, in Code Snippet~\ref{snippet:preCondition}, the programmer has written sufficient proof annotations to prove correctness for the \texttt{Borrow} protocol transition, whereas no proof annotation is added for the \texttt{Return} transition.
The new lemma generated for the \texttt{Borrow} transition will pass automatically using the original proof written by the developer, while the second lemma will fail.
This helps \sys to focus on transitions that fail and only explore the cases that need further proof.
% In case the current proof is sufficient for proving one of the cases and the proof is only failing due to another case, keeping the original proof helps \sys to only conquer the cases that need further proof.

Upon finding the proof for each of these cases, \sys replaces the body of the original lemma with invocations of the newly generated lemmas under the condition of each disjunct.
For instance, \texttt{NextPreservesSafety} lemma body in Code Snippet~\ref{snippet:preCondition} will be replaced as follows:
\begin{lstlisting}[label=snippet:originalProofReplacement, basicstyle=\small\tt,escapechar=$]
  if (exists ...) {
    NextPreservesSafetyDisjunct1(s, s');
  } else if (exists ...) {
    NextPreservesSafetyDisjunct2(s, s');
  }
\end{lstlisting}

% \subsection{Post-condition Expander}

\begin{lstlisting}[label=snippet:postCondition,caption=Removing a post-condition may not actually reduce the number of proof obligations, basicstyle=\small\tt,
float=t!,escapechar=$, numbers=left]
lemma SafetyProof()
  ensures InitIsSafe()
  ensures NextMaintainsSafety()
{
  assert InitIsSafe() by {
    // more proof annotations
  }
  assert NextMaintainsSafety() by {
    // more proof annotations
  }
}
\end{lstlisting}

% Similarly, \sys tries to break the post-conditions of a lemma when there is a conjunct in the post-conditions of a lemma.
\textbf{Sharding Post-conditions:}
Similar to disjuncts in pre-conditions, \sys shards conjuncts in post-conditions and generates a lemma for each shard.
It is common for theorem provers to use {\em short-circuiting} for multiple post-conditions, where each conjunct assumes previous conjuncts are true.
As a result, \sys generates the lemma for the first conjunct with only this conjunct as post-condition.
However, the lemma for subsequent conjuncts includes all previous conjuncts as pre-conditions (and, of course, the current conjunct as a post-condition).
% For instance, the generated lemmas for each conjunct in post-conditions of the \texttt{SafetyProof} lemma in Code Snippet~\ref{snippet:postCondition} are as follows:
% \begin{lstlisting}[label=snippet:ifBranch, basicstyle=\small\tt,escapechar=$]
% lemma FirstConjunct()
%   ensures InitIsSafe()
% lemma SecondConjunct()
%   requires InitIsSafe()
%   ensures NextMaintainsSafety()
% \end{lstlisting}

\manos{This is a big subsection. Need bf headers.}
\armin{break the following sentence to smaller parts.}
Unlike pre-conditions, where keeping the original proof is always useful, breaking post-conditions into multiple conjuncts requires additional care.
Keeping the original proof untouched when sharding post-conditions may result in the same number of proof obligations in the generated lemmas.
Consider the lemma in Code Snippet~\ref{snippet:postCondition} where the programmer has asserted both conjuncts in the body of this lemma.
Due to these assertions, \sys still needs to synthesize the proof for both conjuncts of this lemma even after proof sharding.
% Hence, even if we remove all post-conditions of this lemma (no proof obligations), theorem prover will still fail to prove correctness of this lemma since the theorem prover need additional proof annotations to prove these two assertions.

A naive solution to this problem is to remove the body of the lemma and try to find the proof of each conjunct's lemma from scratch.
However, this solution disallows \sys to leverage\manos{"limit \sys to reuse" means this will force \sys to reuse. I don't think that's what you meant.}\armin{fixed} the manually written proof by the developer for the already proven correct conjuncts.

Our goal is to force the theorem prover to only focus on the particular conjunct we want to prove in each shard while skipping the proof for other conjuncts.
To reuse the current proof of the lemma, \sys wraps the original proof in an \texttt{if} statement to bypass the irrelevant parts of the original proof.
\sys adds the negation of the conjunct we want to prove in each lemma as the condition for the \texttt{if} statement to construct a proof by contradiction.
For instance, in the \texttt{SafetyProof} example above, \sys adds the following if statement for the second conjunct:
\begin{lstlisting}[label=snippet:secondConjunct, basicstyle=\small\tt,escapechar=$]
lemma SecondConjunct()
  requires InitIsSafe()
  ensures NextMaintainsSafety()
{
  if (!NextMaintainsSafety()) {
    // original proof
  }
}
\end{lstlisting}
With this addition, when the theorem prover checks the corresponding lemma for conjunct \texttt{C}, it can now reuse the manually written proof annotations by the programmer.
If the theorem prover proves \texttt{C} at any point of the original proof inside the \texttt{if} statement, it can conclude false, and hence, by contradiction, prove that \texttt{C} is true and pass the proof for this shard.

Note that theorem provers can prove any statement after proving false; as a result, even if the proof for another conjunct is incomplete, the proof for this conjunct will still pass.
In other words, the focus of the proof is now only on the conjunct at hand.\manos{Not sure I followed your reasoning here.}

% \subsection{Definition Expander}
% In some cases, a disjunct in a pre-condition or a conjunct in a post-condition lies within the definition of another function in the pre-condition.
% For instance, the \texttt{Next} predicate in the book shelf example in Code Snippet~\ref{snippet:bookShelfExample} contains a disjunct.
% As a result, if this predicate appears in the pre-condition of a lemma, we need to first expand the definition of the predicate and next use the Pre-condition Expander Lemma Transformer.

% The Definition Expander first determines if multiple disjuncts/conjuncts exists in a predicate in the pre-/post-condition of a lemma or not.
% And in that case, it expands the predicate definition to enable the Pre-/Post-condition Expander lemma transformers to break down the proof even further.

\subsection{Quantifier Elimination Transformer}
Expanding quantifiers is one of the common approaches developers use to reason about program correctness.
As we demonstrated in Section~\ref{sec:motiv}, invoking the necessary trigger to complete the proof is sometimes impossible without introducing new variables into the program scope.
The Quantifier Elimination Lemma Transformer introduces a new variable into the program scope using the existential/universal quantifiers in pre-/post-conditions, respectively.
This enables the exploration lemma transformers to further explore the proof search space using the newly introduced variables.\manos{Is there an example we can point to? It's a bit theoretical otherwise.}
\section{Proving Lemmas at Scale}
To implement \sys, we extend the Dafny~\cite{leino2010dafny} programming language.
We utilize Dafny's Abstract Syntax Tree (AST) to analyze the program and explore the proof search space.
Each lemma transformer designed in \sys adds proof annotations to a lemma and then queries the theorem prover backend.
Upon finding the correct proof for a lemma, \sys displays the complete proof to the user.
Note that since the output of \sys is a Dafny proof, which can be verified using an unmodified version of Dafny, \sys does not increase the trusted computing base (TCB).

\sys generates a large number of lemmas that need to be checked by the theorem prover.
Using a single theorem prover instance to check these lemmas in sequence is neither efficient nor scalable.
\manos{Sure it is. Do you mean not efficient? Or not scalable?}\armin{yes. both}
To check these generated lemmas at scale, we implement a parallel and scalable theorem prover backend in C++.
We run an instance of this backend on each machine of a cluster of nodes.
Each instance can utilize all cores in the machine to check the correctness of lemmas in parallel.
This backend plays a key role in allowing \sys to shard and explore the proof search space effectively and quickly at scale.
\sys sends verification requests to this backend using Google remote procedure calls (gRPC)~\cite{grpc}.
% This theorem prover backend is implemented in 1078 lines of code.
\sys uses a round-robin scheduler to send each verification request to one of the nodes in the backend cluster.
Table~\ref{tabular:linesOfCode} shows the number of lines of code (LoC) for each component of \sys.

% Based on the result of each query, \sys shards and explores the proof search space.
% \sys combines the results of verification queries from all theorem prover backends.

\begin{table}[t!]
\centering
\caption{Lines of code for each component of \sys}
\begin{tabular}{|c|c|c|}
\hline
                            & Language & LoC  \\
\hline
Theorem Prover Backend      & C++      & 1078 \\
\hline
Lemma Transformers          & C\#      & 2373 \\
\hline
\exprGenerator              & C\#      & 2549 \\
\hline
Rest of \sys                & C\#      & 3317 \\
\hline
\end{tabular}
\label{tabular:linesOfCode}
\end{table}

\begin{table*}[t!]
\centering
\caption{List of benchmarks. For each benchmark, the first four columns show the protocol and proof sizes, number of lemmas, and number of single-hole experiments. The last column shows the number of single-hole experiments in which \sys successfully finds the missing proof.}
\begin{tabular}{|l|l|l|l|l|l|}
\hline
Name                                                & Protocol Size (LoC) & Proof Size (LoC) & \# Lemmas &  \begin{tabular}[c]{@{}l@{}}\# Single-Hole \\ Experiments\end{tabular} & Success Rate \\ \hline
Library                                             & 49                  & 16               & 1         & 2        & 2 (100\%)        \\ \hline
Leader Election                                     & 99                  & 47               & 1         & 4        & 4 (100\%)       \\ \hline
Single-Server Lock Service                          & 72                  & 9                & 1         & 1        & 1 (100\%)       \\ \hline
Echo Server~\cite{Zhang2025Singh}                   & 151                 & 1                & 1         & 1        & 1 (100\%)       \\ \hline
Paxos~\cite{lamport2001paxos}                       & 410                 & 476              & 17        & 31       & 27 (87.1\%)      \\ \hline
Reduce~\cite{Zhang2025Singh}                        & 119                 & 28               & 2         & 4        & 4 (100\%)       \\ \hline
Raft Leader Election~\cite{Ongaro2014Ousterhout}    & 154                 & 56               & 2         & 7        & 1 (14.3\%)       \\ \hline
Ring Leader Election~\cite{Chang1979Roberts}        & 77                  & 81               & 3         & 8        & 8 (100\%)       \\ \hline
Simplified Leader Election~\cite{Taube2018Losa}     & 129                 & 71               & 3         & 11       & 9 (81.8\%)       \\ \hline
Two-Phase Commit~\cite{Zhang2025Singh}              & 202                 & 93               & 6         & 16       & 15 (93.8\%)      \\ \hline
Three-Phase Commit~\cite{Skeen1983Stonebraker}      & 247                 & 105              & 6         & 18       & 14 (77.8\%)      \\ \hline
\end{tabular}
\label{tabular:benchmarks}
\end{table*}

\section{Evaluation}
We use ten CloudLab~\cite{Duplyakin+:ATC19} c220g2 nodes to run the theorem prover backend.
Each node has two Intel Xeon E5-2660 processors with 10-core CPUs. Jointly, these nodes can verify 200 lemmas in parallel (one per physical core).
We run \sys on a separate CloudLab c220g2 node, which sends verification requests through gRPC~\cite{grpc} to the other ten c220g2 nodes running the theorem prover backend.
% \tpas plays a key role in allowing \sys to check the generated lemmas in parallel quickly.

% \begin{table*}[t!]
% \centering
% \caption{List of Benchmarks}
% \begin{tabular}{|l|l|l|l|l|l|}
% \hline
% Name                                             & Protocol       & Proof       & \# Lemmas \\
%                                                  & (LoC)          & (LoC)       & \\ \hline
% Library                                        & 49             & 16          & \\ \hline
% Leader Election                                  & 99             & 47          & \\ \hline
% Single-Server Lock Service                       & 72             & 9           & \\ \hline
% Echo Server~\cite{Zhang2025Singh}                & 151            & 1           & \\ \hline
% Paxos~\cite{Zhang2025Singh}                      & 410            & 476         & \\ \hline
% Reduce~\cite{Zhang2025Singh}                     & 119            & 28          & \\ \hline
% Raft~\cite{Zhang2025Singh}                       & 154            & 56          & \\ \hline
% Ring Leader Election~\cite{Zhang2025Singh}       & 77             & 81          & \\ \hline
% Simplified Leader Election~\cite{Zhang2025Singh} & 129            & 71          & \\ \hline
% Two-Phase Commit~\cite{Zhang2025Singh}           & 202            & 93          & \\ \hline
% Three-Phase Commit~\cite{Zhang2025Singh}         & 247            & 105         & \\ \hline
% \end{tabular}
% \label{tabular:benchmarks}
% \end{table*}

We evaluate \sys on lemmas that theorem provers fail to prove automatically and require extra proof annotations.
% To evaluate \sys, we use it to synthesize the proof of lemmas where theorem provers fail to prove without extra proof annotations\manos{weird phrasing}.
We apply \sys to three in-house protocols and all protocols used in the evaluation of Basilisk~\cite{Zhang2025Singh}, except for those simple enough to not require a manual proof.
% We also evaluated \sys on a variety of in-house protocols.
Table~\ref{tabular:benchmarks} shows the list of benchmarks we use to evaluate \sys.

Programmers usually publish their code once the code is formally verified, and they do not publish the history of how they reached the final proof.
% We evaluate \sys on two category of benchmarks. First, we remove proof annotations from a complete proof and evaluate if \sys can find the missing proof or not. 
As a result, we were not able to evaluate \sys on failing proofs that programmers have encountered during proof debugging.
Instead, we evaluate \sys starting from the complete proof that programmers have published, then remove one or more lines from the proof and check whether \sys can complete the proof after the removal.
We only consider removing proof annotations where the proof fails after their removal due to theorem prover failure, and not due to a parse error.
We call a single removal of a trigger, function invocation, etc., a {\em hole}.
We evaluate \sys by removing one (\S\ref{sec:eval_single_hole}) or multiple (\S\ref{sec:eval_multi_hole}) holes and check if it can find the missing proof.
We call each of these removals one \textit{experiment}.

% A second set of benchmarks we evaluate \sys on are cases where upgrading Dafny or Z3 results into having a failing proof. Due to flakiness of proofs, programmers sometimes need to manually fix the proofs that were done before on an older version of Dafny again. In such cases, we use \sys to fix the proofs and find the missing annotations after upgrading Dafny to a later version automatically.
% We aim to show how much \sys is able to help the programmer in proof debugging

% The number of experiments we executed for each benchmark is shown in Table~\ref{tabular:benchmarks}.

Our goal is to determine whether \sys can help developers find the proof of lemmas during the proof debugging phase.
We answer the following questions to establish this:
\squishlist
\item How effective is \sys in finding the proof of failing lemmas during proof development? (\S\ref{sec:eval_single_hole}, \S\ref{sec:eval_multi_hole})
\item Can \sys find the proof of lemmas from scratch? (\S\ref{sec:eval_scratch})
% \item What is \sys success rate in finding missing proof annotations at each complexity level? (\S\ref{sec:eval_hole_complexity})
% \item How long does it take for \sys to find the correct proof? (\S\ref{sec:eval_exec})
% \item How many expressions \sys generates at each complexity level? (\S\ref{sec:eval_expr_complexity})
\item How effectively does the theorem-prover backend leverage a cluster of nodes to scale \sys? (\S\ref{sec:eval_scalability})\manos{I'm not sure it's improving the scalability, per se. It's leveraging the parallelism in \sys's workload to speed up verification. Not sure exactly how to phrase this.}
\squishend

\begin{figure}[t!]
\centering
\includegraphics[width=\columnwidth]{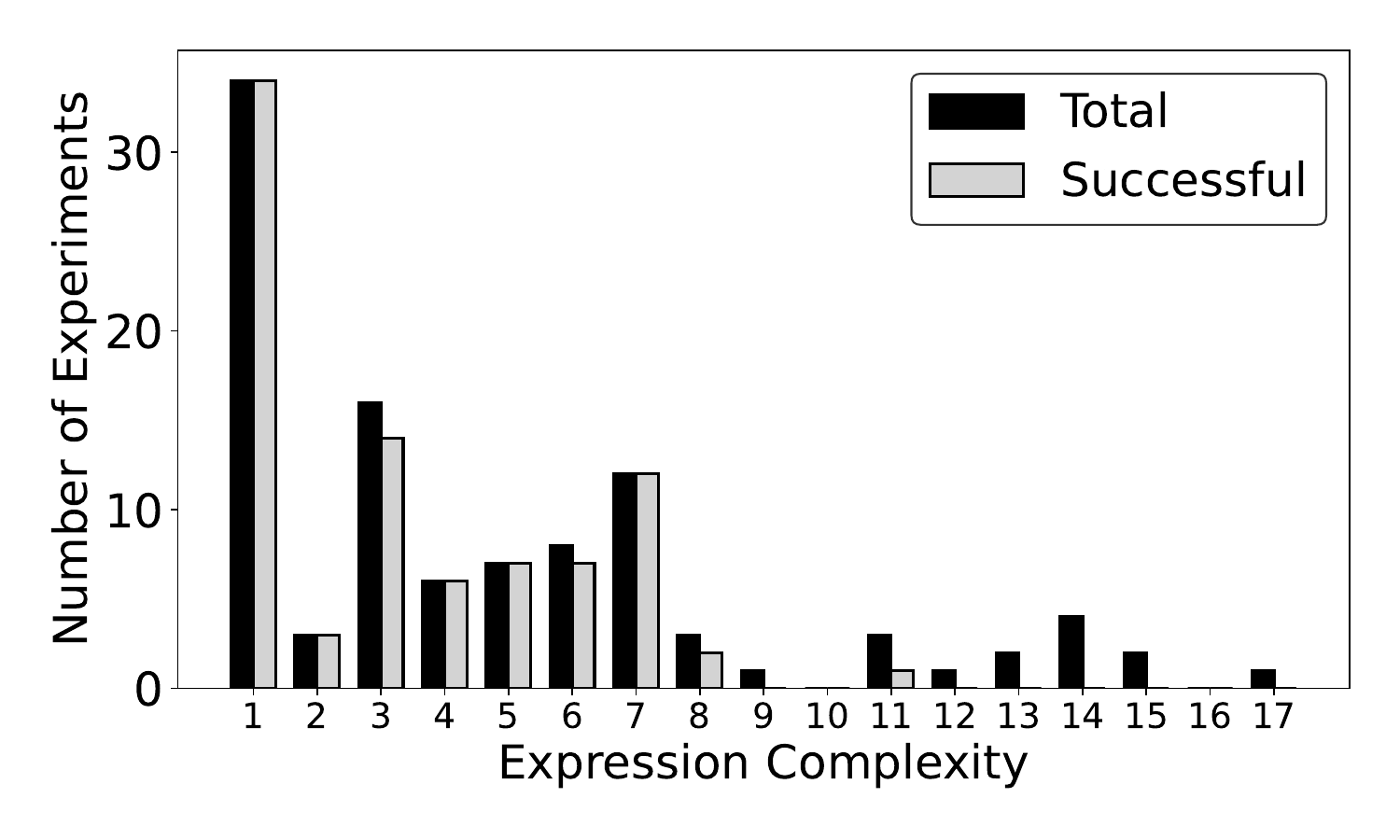}
\caption{Number of single-hole experiments per expression complexity and number of cases where \sys finds the proof successfully.}
\Description{Number of single-hole experiments per expression complexity and number of cases where \sys finds the proof successfully.}
\vspace{-5pt}
\label{fig:successPerComplexity}
\end{figure}

\begin{figure}[t!]
\centering
\includegraphics[width=\columnwidth]{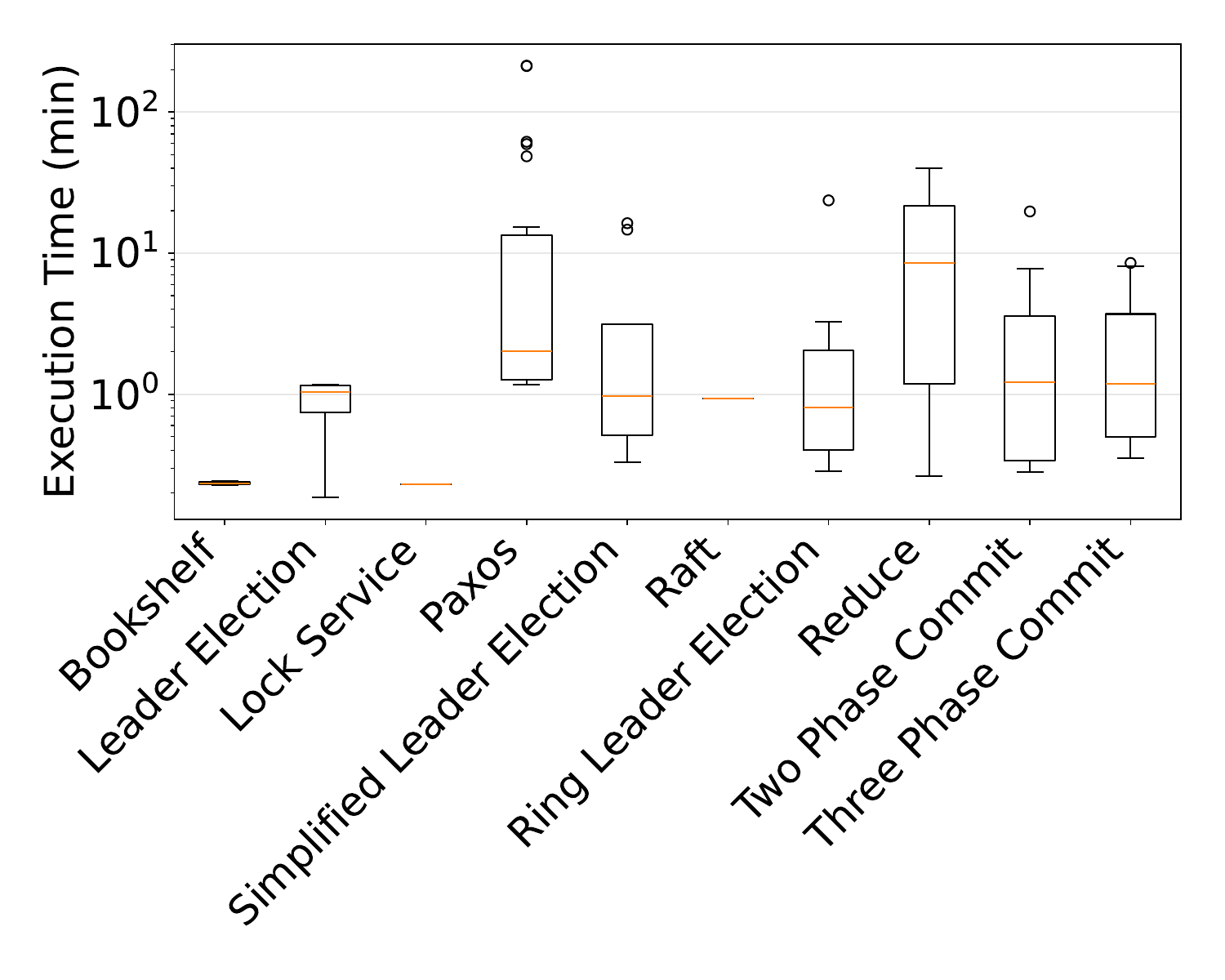}
\caption{\sys execution time when successful in finding the correct proof in single-hole experiments.}
\Description{\sys execution time when successful in finding the correct proof in single-hole experiments.}
\vspace{-5pt}
\label{fig:execTime}
\end{figure}

% \begin{table}[t!]
% \centering
% \caption{Number of Single-Hole Experiments}
% \begin{tabular}{|l|l|l|}
% \hline
% Name                       & \# Experiments & Success \\ \hline
% Library                  & 2              & 2        \\ \hline
% Leader Election            & 4              & 4       \\ \hline
% Single-Server Lock Service & 1              & 1       \\ \hline
% Echo Server~\cite{Zhang2025Singh}                & 1              & 1      \\ \hline
% Paxos~\cite{Zhang2025Singh}                      & 32               & 26       \\ \hline
% Reduce~\cite{Zhang2025Singh}                     & 4               & 4         \\ \hline
% Raft~\cite{Zhang2025Singh}                       & 7              & 1        \\ \hline
% Ring Leader Election~\cite{Zhang2025Singh}       & 8             & 8        \\ \hline
% Simplified Leader Election & 11             & 10       \\ \hline
% Two-Phase Commit~\cite{Zhang2025Singh}           & 16             & 12        \\ \hline
% Three-Phase Commit~\cite{Zhang2025Singh}         & 18             & 12        \\ \hline
% \end{tabular}
% \label{tabular:experiments}
% \end{table}

\subsection{\sys Finding Single Holes}\label{sec:eval_single_hole}
\bubble{\sys can find the proof in 86 cases.}
First, we evaluate the effectiveness of \sys on proofs that have a single hole. In total, we perform 103 \textit{experiments}. The number of these experiments for each benchmark is listed in Table~\ref{tabular:benchmarks}.
\sys is able to find the proof after the removal of a proof annotation from the complete proof in 86 cases overall.

\textbf{Complexity of Single-Hole Experiments:}
Figure~\ref{fig:successPerComplexity} shows the number of single-hole experiments we use to evaluate \sys.
For each expression-complexity level in the missing proof, the figure shows both the total number of experiments and the number of experiments in which \sys successfully finds the correct proof.\manos{Too long sentence. Split in two. Also, the "per complexity" phrasing is weird.}
% This figure shows \sys can find most of the missing proof annotation at expression complexities of below seven.
In the 17 cases where \sys fails to find the correct proof, the main contributing reasons are 1) high complexity of the missing proof annotation and 2) unsupported language features such as templates and set comprehensions.
% \sys is able to find the proof in all experiments where the missing proof annotation is included in the grammar of expressions \sys generates. \armin{not sure if we need this sentence.}\manos{I'm more worried about the sentence not being true. Above, we said that one reason for not finding an annotation is because it is too long. That seems to contradict this sentence.}

% \bubble{In cases that \sys fails, it is due to (1) high complexity or (2) unsupported language features not generated in \sys language.}

\bubble{\sys execution time depends on the complexity of missing proof.}
Figure~\ref{fig:execTime} shows the execution time of \sys when it successfully finds the missing proof for the \textit{experiments} described above.
Apart from a few edge cases in the Paxos protocol, \sys can successfully find the missing proof within 20 minutes.
This shows that \sys can help developers find the missing proof in a reasonable amount of time and is scalable even for complex proofs such as the Paxos consensus protocol.

% The blue line with circle markers represents the cases where \sys successfully finds the proof whereas the red line with square markers shows the cases where \sys is unsuccessful to find the correct proof.
% \sys can find the missing proof in range of a minute to 2 hours in the worst case for one of the experiments in Paxos benchmark.

% Figure~\ref{fig:execTime}(b) shows the execution time of experiments that \sys does not find the missing proof successfully.
% This figure depicts how long programmers need to wait on \sys response to determine \sys cannot help in finding the complete proof and a manual proof is required.
% Note that even in these cases, \sys provides a partial proof for parts that it could find the missing proof and programmers can use this as guidance on which exact parts of their protocol needs extra proof annotations.

% The experiments in this figure are sorted based on execution time of \sys from left to right, which is correlated with the complexity of the missing proof.

\bubble{A variety of lemma transformers contribute to find the missing proof of a lemma.}
\textbf{Effectiveness of lemma transformers:}
In each experiment, one or more lemma transformers contributed to finding the correct proof.
Table~\ref{tabular:conqueror} shows how many experiments each transformer type helped solve.\manos{convoluted syntax. Maybe avoid passive voice, if possible.}

\begin{table}[t!]
\centering
\caption{Number of experiments each lemma transformer helped solve}
\begin{tabular}{|c|c|}
\hline
Name                       & \# Experiments    \\ \hline
Trigger Synthesizer              & 9   \\ \hline
Lemma Invocation                 & 42  \\ \hline
Reveal Opaque Definitions        & 38  \\ \hline
Expanding Pre-/Post-conditions   & 6   \\ \hline
Quantifier Elimination           & 5   \\ \hline
\end{tabular}
\label{tabular:conqueror}
\vspace{-10pt}
\end{table}

% In the XXX experiments that \sys fails to find the correct proof for, the contributing reasons are 1) high complexity of the missing proof annotation and 2) unsupported language features such as templates.
% \sys is able to find the proof of experiments where the missing proof annotation is included in the grammar of expressions \sys generates.

\sys is not limited to finding proofs where only one expression is missing.
In the next two sections, we show that the main limiting factor in automatically finding the proof of lemmas with more than one missing proof annotation is the complexity of the missing proof in each \textit{shard}. We show that \sys can complete the proof even with multiple holes in the same lemma, as long as these holes do not fall within the same shard.

\subsection{Multi-hole Experiments}\label{sec:eval_multi_hole}
\bubble{\sys is not limited to find only single holes.}
We also apply \sys to lemmas where the programmer needs to add more than one proof annotation to fix the proof.
In these experiments, we only include cases where \sys is able to find the proof when only one of the missing proof annotations is removed.
We exclude cases where \sys cannot solve the corresponding single-hole experiment.\manos{This previous sentence is WAY too complex.}
Out of 42 multi-hole experiments, \sys can find the correct proof in 32 experiments.

\sys is effective in finding the proof in all cases where, after sharding, the remaining proof annotation to be synthesized has low expression complexity.
On the other hand, if sharding is not successful in breaking down the proof and the missing proof annotation requires, say, multiple lemma invocations, \sys fails to find the missing proof.
Out of the ten cases that \sys does not find the proof, seven of them require two lemma invocations to find the proof.
\sys intentionally does not synthesize multiple lemma invocations because this leads to search-space explosion.
% There are two factors that prevent \sys to find the proof in the remaining XXX cases.

% \bubble{The main bottleneck for \sys is complexity and expressivity of the missing proof.}
% The main challenge for \sys in finding the correct proof comes from the complexity and expressivity of the missing proof. 
% These experiments 

% \subsection{Proof repair experiments}\label{sec:eval_repair}

\subsection{Finding Proofs from Scratch}\label{sec:eval_scratch}
Fully automating proof search for a lemma is an undecidable problem and naturally is unlikely to work in all cases.
\sys lays the foundation for controlled exploration of proof annotations and provides the groundwork for subsequent development of scalable proof synthesis.\manos{I see what you wanted to do here, but this paragraph feels unnaturally out of place and doesn't smoothly make the desired point. It would be better to FIRST make the point that finding a proof from scratch is really hard and THEN present the results in the tone of "Despite this difficulty, \sys can still find the proof from scratch in 8 out of 49 experiments.."}

Despite this difficulty, we evaluate \sys on the benchmarks listed in Table~\ref{tabular:benchmarks} after removing {\em all} manually written proof annotations in each lemma.
We run one experiment for each lemma in these benchmarks.
Among these experiments, \sys is able to find the proof from scratch for eight out of 49 experiments.

\begin{lstlisting}[label=snippet:ringProof,caption=Summary of the manually written proof for the Ring leader election protocol from Basilisk~\cite{Zhang2025Singh}, basicstyle=\small\tt, float=t!,escapechar=$, numbers=left]
lemma MsgInvsImplyChordDominates(
    c: Constants, v: Variables) ... {
  forall src:nat, dst:nat, mid:nat | 
    ...
    ensures
    c.hosts[mid].hostId<c.hosts[src].hostId
  {
    reveal_ValidHistory();
    HostReceiveSkolemization(
        c, v, |v.history|-1, dst);
    if Predecessor(|c.hosts|, dst) != mid {
      MidMustHaveSentSrcHostId(
        c, v, src, mid, dst);
    }
  }
}
lemma MidMustHaveSentSrcHostId(
  c: Constants, v: Variables, 
  src: nat, mid: nat, dst: nat) ... {
  LemmaSentNotMyIdImpliesReceivedId(c, v);
  var n := |c.hosts|;
  if mid == Predecessor(n, dst) {
    reveal_ValidHistory();
    HostReceiveSkolemization(
        c, v, |v.history|-1, dst);
  } else {
    MidMustHaveSentSrcHostId(
        c, v, src, Successor(n, mid), dst);
  }
}
lemma LemmaSentNotMyIdImpliesReceivedId(
    c: Constants, v: Variables) ... {
  forall msg | 
    msg in v.network.sentMsgs &&
    msg.val != c.hosts[msg.src].hostId 
  ensures 
    Msg(msg.val, 
      Predecessor(|c.hosts|, msg.src))
    in v.network.sentMsgs
  {
    SendMsgSkolemization(c, v, msg);
  }
}
\end{lstlisting}

Focusing on the ring leader election protocol proof from Basilisk~\cite{Zhang2025Singh}, its safety proof consists of three main lemmas.
Code Snippet~\ref{snippet:ringProof} summarizes the manually written proof for these three lemmas.
These three lemmas are intertwined, as each of the first two lemmas invokes the next one\manos{Each? Even the last one? Can't be right or they'd be recursive}.
Moreover, the second lemma contains a recursive lemma invocation to itself.

In the manually written proof for the first lemma, the programmer first introduces new variables into the scope using a \texttt{forall} quantifier, followed by revealing the opaque predicate, \texttt{ValidHistory}.
Next, the programmer invokes two helper lemmas under some condition.
The second lemma invokes another helper lemma under a different condition and otherwise invokes a recursive call to itself.
Finally, the last lemma introduces\manos{includes expanding? that's weird phrasing} the \texttt{msg} variable into the scope, followed by a helper lemma invocation.
% As depicted in the single-hole experiments, \sys is able to find the proof when 

As depicted in the single-hole experiments, \sys is able to find the proof after all valid single-line removals from this proof.
\sys is also able to find the proof for the first and last lemmas if the lemma body is empty.
% The proof that \sys finds for these two lemmas includes removing\manos{again, "includes removing... and then invoking". That is very odd syntax.} the \texttt{forall} quantifier, and then invoking another lemma.
\sys finds the proof for the first lemma in 148 seconds while it finds the proof for the last lemma in 75 seconds.
% Note that \sys finds a simpler proof for the first lemma than the manually written proof.

On the other hand, \sys does not find the proof for the second lemma after two hours of execution time and checking 235 thousand lemmas using the theorem prover.
\sys supports recursive lemma invocations and that is not the reason \sys fails to find the proof for this lemma.
However, \sys does not synthesize \texttt{if} conditions such as line 22 because these do not fall into the proof annotation patterns \sys synthesizes.\manos{Should we say why it does not synthesize those?}

It is important to note that \sys does not synthesize helper lemmas on its own.
As a result, if the last two lemmas' signatures are not available in the code, \sys would not be able to find the proof for the first lemma from scratch since the first lemma invokes the second lemma.
The goal of \sys is to lay the foundation for scaling proof synthesis for the last steps of formal verification, where only a few proof annotations are left to find the complete proof of a lemma after proof sharding.
The last steps of verification usually involve finding triggers, invoking other helper lemmas, or revealing opaque definitions, all of which are explored during the proof exploration phase of \sys.

% \subsection{\sys Execution time}\label{sec:eval_exec}

% \begin{figure}[t!]
% \centering
% \includegraphics[width=\columnwidth]{figures/trigger_lemma_per_depth.pdf}
% \caption{Number of triggers and lemma invocations generated during the conquer phase in \sys per expression complexity (i.e., number of operators and operands).}
% \Description{Number of triggers and lemma invocations generated during the conquer phase in \sys per expression complexity (i.e., number of operators and operands).}
% \label{fig:sensExprComplexity}
% \end{figure}

% \subsection{Expressions Complexity}\label{sec:eval_expr_complexity}
% During the conquer phase, \sys generates expressions that match a trigger or a lemma invocation based on a maximum complexity at each step.
% It starts from expressions with lowest complexity and increases the complexity at each step if a proof is not found.
% Figure~\ref{fig:sensExprComplexity} shows the number of generated triggers and lemma invocations when \sys synthesizes the proof for the second lemma in Code Snippet~\ref{snippet:ringProof} from scratch.
% As mentioned in \S\ref{sec:eval_scratch}, \sys does not find the proof for this lemma from scratch after two hours of execution time.

% The number of generated triggers in this experiment varies from 5 to 541 

\subsection{\sys Scalability}\label{sec:eval_scalability}
% To evaluate how effective is the theorem-prover backend in scaling the execution of \sys, we show how the execution time of \sys to find a single hole in the ring leader election benchmark in Code Snippet~\ref{snippet:ringProof} varies when we change the number of nodes used in theorem-prover backend from one to ten nodes.
To evaluate the scalability of the theorem-prover backend, we measure how the runtime of \sys changes as we increase the number of backend nodes from one to ten.
% for finding a single hole in the ring leader election benchmark (Code Snippet~\ref{snippet:ringProof})

\begin{figure}[t!]
\centering
\includegraphics[width=\columnwidth]{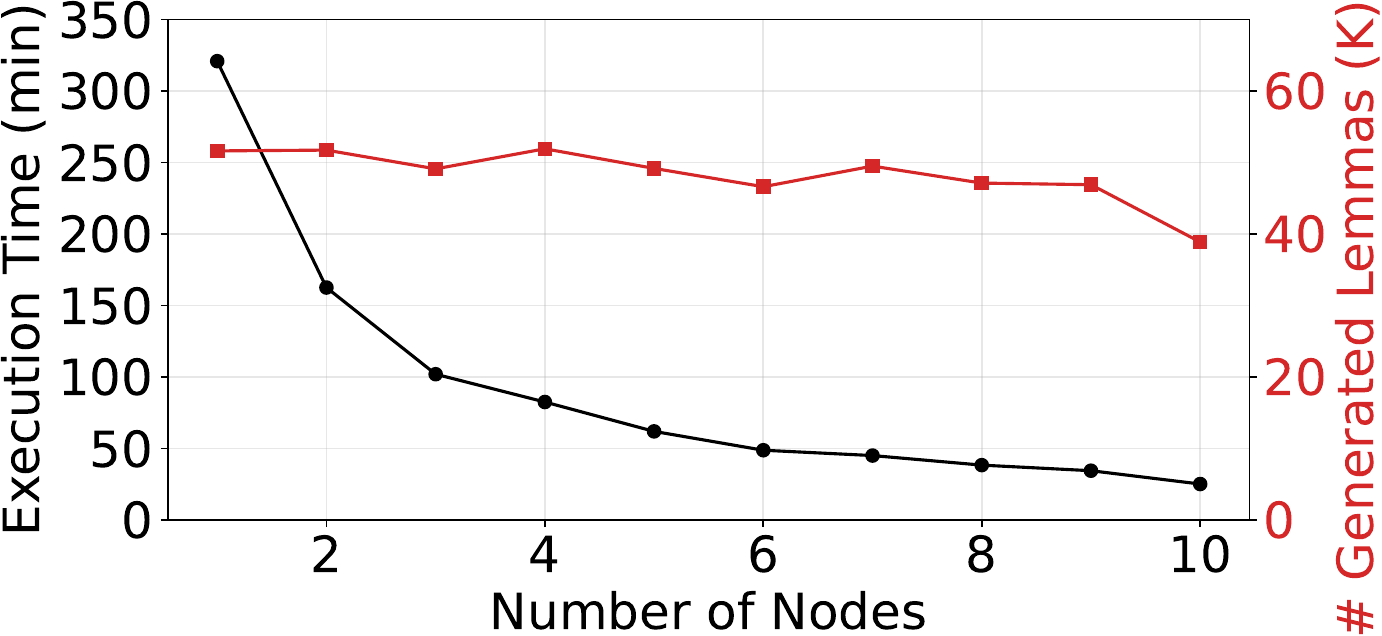}
\caption{\sys execution time when changing the number of nodes in the theorem-prover backend.}
\Description{\sys execution time when changing the number of nodes in the theorem-prover backend.}
\label{fig:eval_scale}
\vspace{-5pt}
\end{figure}

For this analysis, we remove the lemma invocations in lines 27 and 28 in Code Snippet~\ref{snippet:ringProof} and analyze the execution time of \sys by changing the number of available nodes in the backend.
% Figure~\ref{fig:eval_scale} shows how \sys execution time  changes when changing the number of nodes from one to ten.
Figure~\ref{fig:eval_scale} shows the execution time of \sys in this analysis.
% Using multiple nodes, the theorem-prover backend reduces the time required to find the proof from more than 6 hours when only one node with 20 physical cores is available to 25 minutes with 10 nodes.
Increasing the number of backend nodes significantly reduces the proof search time. \sys finds the missing proof annotation with a single node (20 physical cores) in over six hours, while using ten nodes reduces the execution time of \sys to under 25 minutes.

The number of lemmas generated by \sys to find the proof of this example is shown in red in Figure~\ref{fig:eval_scale}.
The total number of lemmas generated to find the proof in this example ranges from 52K with one node to 38K with 10 nodes.
This number decreases if the correct proof is found more quickly, as \sys keeps generating lemmas with higher complexity until it can find the proof.

% In the next sections, we show in how many cases \sys can find the proof after one or few removals. And we also show in how many cases \sys can find the proof without any guidance from the developer from scratch.

\section{Discussion}
% \subsection{Trusted Computing Base}
% Although our changes are directly done in Dafny code base, we do not increase the trusted computing base (TCB) of formal verification.
% This is due to the fact that \sys only generate proofs that programmers can then use to run the theorem prover independently.

% \subsection{Multiple Failing Lemmas}
% \armin{move this to 3 or 4 as its important}
% \sys focus is only on the body of lemmas and does not synthesize or change signature (pre- and post-condition) of lemmas.
% As a result, lemmas can be independently synthesized without impacting each other.
% Hence, if multiple lemmas in a proof do not verify automatically via the theorem prover, \sys can synthesize the proof for each lemma independently.

% \subsection{\sys workflow upon failure to find the missing proof}

\subsection{Timeouts}
While debugging proofs, programmers often encounter cases where the theorem prover times out without giving any insight to the programmer.
In such cases, programmers cannot rely on the theorem prover's output to know which proof obligation requires further proof annotations to prove correctness.
This makes debugging proofs even more challenging and is a source of frustration for both newcomers and experts in formal verification~\cite{Leino2016Pit-Claudel}.
The divide principle in \sys helps developers break down their proof when encountering timeouts, and eventually find the proof for individual parts of their program independently.

\subsection{Trigger-based Programming Languages}
The ideas presented in this paper are not limited to the Dafny~\cite{leino2010dafny} programming language.
Other formal verification languages that use Z3~\cite{Moura2008Bjonrner} as their underlying theorem prover, such as F*~\cite{Swamy2016Hrictcu} or Verus~\cite{lattuada2024verus}, and similarly use triggers to instantiate quantifiers can leverage the techniques presented in this paper to synthesize trigger invocations.

\section{Related Work}

\textbf{Distributed systems verification} has been the focus of a variety of work in the past decade.
Ironfleet~\cite{Hawblitzel2015Howell} and Verdi~\cite{Wilcox2015Woos} pioneered the use of formal verification to verify the correctness of distributed systems.
They demonstrated how to use theorem provers to verify safety and liveness properties of distributed systems. These proofs had to be written manually by developers.
%While formal verification offers highest assurance on protocol correctness, as demonstrated in these works, it comes at a high cost of writing manual proofs.

In particular, developers have to (1) find inductive invariants that hold throughout the execution and (2) prove that these invariants are indeed inductive.
Subsequent work~\cite{padon2016ivy,ma2019i4,Yao2022Tao,Yao2021Tao} introduced automation for finding the inductive invariants of distributed protocols, but only in the restricted and less practical setting of decidable protocols using effectively propositional reasoning (EPR)~\cite{Moura2008Bjorner}.
Zhang~et~al.~\cite{Zhang2024Hance, Zhang2025Singh} remove this requirement and automatically find the inductive invariant for complex distributed protocols written using undecidable logic.

Although these works help find inductive invariants, they still rely on the developer to prove manually that the candidate invariants produced by these tools are indeed inductive. \sys complements these works by providing automation for finding the correct proof using proof sharding.

\textbf{Interactive theorem provers:}
\sys is not the first work to attempt synthesizing proofs of correctness.
A large body of work uses proof synthesis to help programmers when writing proofs using interactive theorem provers such as Rocq~\cite{coq} or Isabelle~\cite{Nikpow2002Wenzel}.
CoqHammer~\cite{Czajka2018Kaliszyk} and SledgeHammer~\cite{Meng2009Paulson} provide automated tactics that programmers may use during proof debugging in Rocq and Isabelle, respectively.
Unlike \sys, these works still keep the developer in the proof-debugging loop depicted in Figure~\ref{fig:proofDebugging}. They do help by making each iteration of the loop faster, but the developer still has to decide which annotation or tactic to add in order to convince the prover.

%After the theorem prover fails to prove a lemma, programmer needs to come up with the next proof annotation or tactic to add to the proof in order to convince the solver.
%\manos{So how do they help? Also, couldn't you say the same about \sys? What distinguishes \sys from these works?}
%\armin{These work only provide tactics/annotations for the developer to invoke manually and see if they succeed. They do not remove the programmer from the proof debugging cycle. It is still up to the developer to invoke these tactics where necessary.}

\textbf{Automated code manipulations:}
Another work related to \sys is ProofPlumber~\cite{Cho2024Zhou}.
In this work, Cho~et~al. provide automated code manipulations at the source-code level that programmers may use to debug their proofs.
Similar to CoqHammer and SledgeHammer, ProofPlumber requires manual intervention by the user upon a proof failure.
The programmer needs to decide what proof annotation to try next after the theorem prover fails to prove correctness and is heavily involved in the cycle of going back and forth with the theorem prover to debug the proof.

\textbf{Large Language Models (LLMs):}
With recent advancements in LLMs, a new line of work has started using LLMs to synthesize proofs of lemmas automatically for both Dafny and Verus~\cite{Yang2025Li, silva2025inferringmultiplehelperdafny, poesia2024dafnyannotatoraiassistedverificationdafny}.
Although these works do not focus on distributed systems verification, we believe that LLMs have significant potential in helping synthesize such proofs. That said, we believe that it is preferable to first explore a new area with conventional methods and predictable solutions to better understand its fundamental requirements and subtleties (e.g., how the number of holes per shard can be the limiting factor to scalability). Enhancing \sys with LLMs remains a promising avenue for future work. 

% programmers still need to manually query the theorem prover or the AI agent to determine what proof annotation to try next after a failure due to incomplete proof by LLM or theorem prover requiring extra proof annotations.\manos{Again, it is not clear how \sys is better than this. We also don't guarantee we'll complete the proof.}
% As future work, this work can be integrated with an AI agent\manos{That's too vague. If you are saying that they are complementary, you should explain why you think that.} to enhance its efficiency in finding correct proofs.

% generating The main focus of these work~\cite{} is to automatically generating proof tactics

% Prior work on proof synthesis has tackled finding proofs for lemmas automatically.

\section{Conclusion}
This paper introduces \sys, a controlled proof-search approach for synthesizing proofs of correctness.
We employ a shard-and-explore technique to break down the proof obligations where possible and then explore the proof search space for each shard independently.
\sys employs a parallel theorem prover backend to scale the verification of large proofs.
% We identify patterns of proof annotations programmers typically use and synthesize proofs using these patterns.
Lastly, we show that \sys can find proofs for complex distributed systems when they require only a few proof annotations in each shard. We show that \sys can find the proof of correctness in 86 out of 103 cases, with execution times ranging from one minute to two hours.

\bibliographystyle{ACM-Reference-Format}
\bibliography{refs}

\end{document}